\documentclass[sigconf]{acmart}

\setcopyright{acmcopyright}
\copyrightyear{2024}
\acmYear{2024}
\setcopyright{acmlicensed}\acmConference[CHI '24]{Proceedings of the CHI Conference on Human Factors in Computing Systems}{May 11--16, 2024}{Honolulu, HI, USA}
\acmBooktitle{Proceedings of the CHI Conference on Human Factors in Computing Systems (CHI '24), May 11--16, 2024, Honolulu, HI, USA}
\acmDOI{10.1145/3613904.3642212}
\acmISBN{979-8-4007-0330-0/24/05}

\AtBeginDocument{%
  \providecommand\BibTeX{{%
    \normalfont B\kern-0.5em{\scshape i\kern-0.25em b}\kern-0.8em\TeX}}}

\def\markup{0}

\if\markup1
\usepackage[normalem]{ulem}
  \definecolor{myblue}{rgb}{0,0,0.75}
  \newcommand{\rv}[1]{{\leavevmode\color{myblue}#1}}
  \newcommand{\st}[1]{{\sout{#1}}}
\else
  \newcommand{\rv}[1]{#1}
\newcommand{\st}[1]{}
\newcommand{\sout}[1]{}
\fi

\newcommand{\sys}[0]{{{\it CoPrompt}}}
\newcommand{\referlink}[0]{{refer}}
\newcommand{\requestlink}[0]{{request}}
\newcommand{\respondlink}[0]{{share}}
\newcommand{\reciprocatelink}[0]{{link}}

\usepackage{hyperref}
\usepackage{wrapfig}
\usepackage{xspace}
\usepackage{float}

\usepackage{tabularx}

\begin{document}

\title[CoPrompt: Collaborative Prompt Engineering]{CoPrompt: Supporting Prompt Sharing and Referring in Collaborative Natural Language Programming}

\author{Li Feng}
\authornote{Both authors contributed equally.}
\orcid{0000-0002-6198-0896}
\affiliation{
  \institution{Computational Media and Arts Thrust}
  \institution{The Hong Kong University of Science and Technology (Guangzhou)}
  \city{Guangzhou}
  \country{China}
}

\author{Ryan Yen}
\authornotemark[1]
\orcid{0000-0001-8212-4100}

\affiliation{
  \institution{School of Computer Science}
  \institution{University of Waterloo}
  \streetaddress{200 University Ave W}
  \city{Waterloo}
  \state{Ontario}
  \country{Canada}
}

\author{Yuzhe You}
\orcid{0009-0004-7830-4239}

\affiliation{
  \institution{School of Computer Science}
  \institution{University of Waterloo}
  \streetaddress{200 University Ave W}
  \city{Waterloo}
  \state{Ontario}
  \country{Canada}
}

\author{Mingming Fan}
\orcid{0000-0002-0356-4712}
\affiliation{%
  \institution{Computational Media and Arts Thrust}
  \institution{The Hong Kong University of Science and Technology (Guangzhou)}
  \city{Guangzhou}
  \country{China}
}

\author{Jian Zhao}
\orcid{0000-0001-5008-4319}

\affiliation{%
  \institution{School of Computer Science}
  \institution{University of Waterloo}
  \streetaddress{200 University Ave W}
  \city{Waterloo}
  \state{Ontario}
  \country{Canada}
}

\author{Zhicong Lu}
\orcid{0000-0002-7761-6351}

\affiliation{%
  \institution{Department of Computer Science}
  \institution{City University of Hong Kong}
  \streetaddress{83 Tat Chee Ave}
  \city{Hong Kong}
  \country{China}
}

\renewcommand{\shortauthors}{Feng and Yen et al.}
\newcommand{\qt}[1]{\textit{``#1''}}
\newcommand{\pqt}[2]{\textit{``#1''}{\,}{\small-#2}}

\begin{abstract}
Natural language (NL) programming has become more approachable due to the powerful code-generation capability of large language models (LLMs). This shift to using NL to program enhances collaborative programming by reducing communication barriers and context-switching among programmers from varying backgrounds. However, programmers may face challenges during prompt engineering in a collaborative setting as they need to actively keep aware of their collaborators' progress and intents. In this paper, we aim to investigate ways to assist programmers’ prompt engineering in a collaborative context. We first conducted a formative study to understand the workflows and challenges of programmers when using NL for collaborative programming. Based on our findings, we implemented a prototype, \sys{}, to support collaborative prompt engineering by providing referring, requesting, sharing, and linking mechanisms. Our user study indicates that \sys{} assists programmers in comprehending collaborators' prompts and building on their collaborators’ work, reducing repetitive updates and communication costs. 
\end{abstract}




\begin{teaserfigure}
  \includegraphics[width=\textwidth]{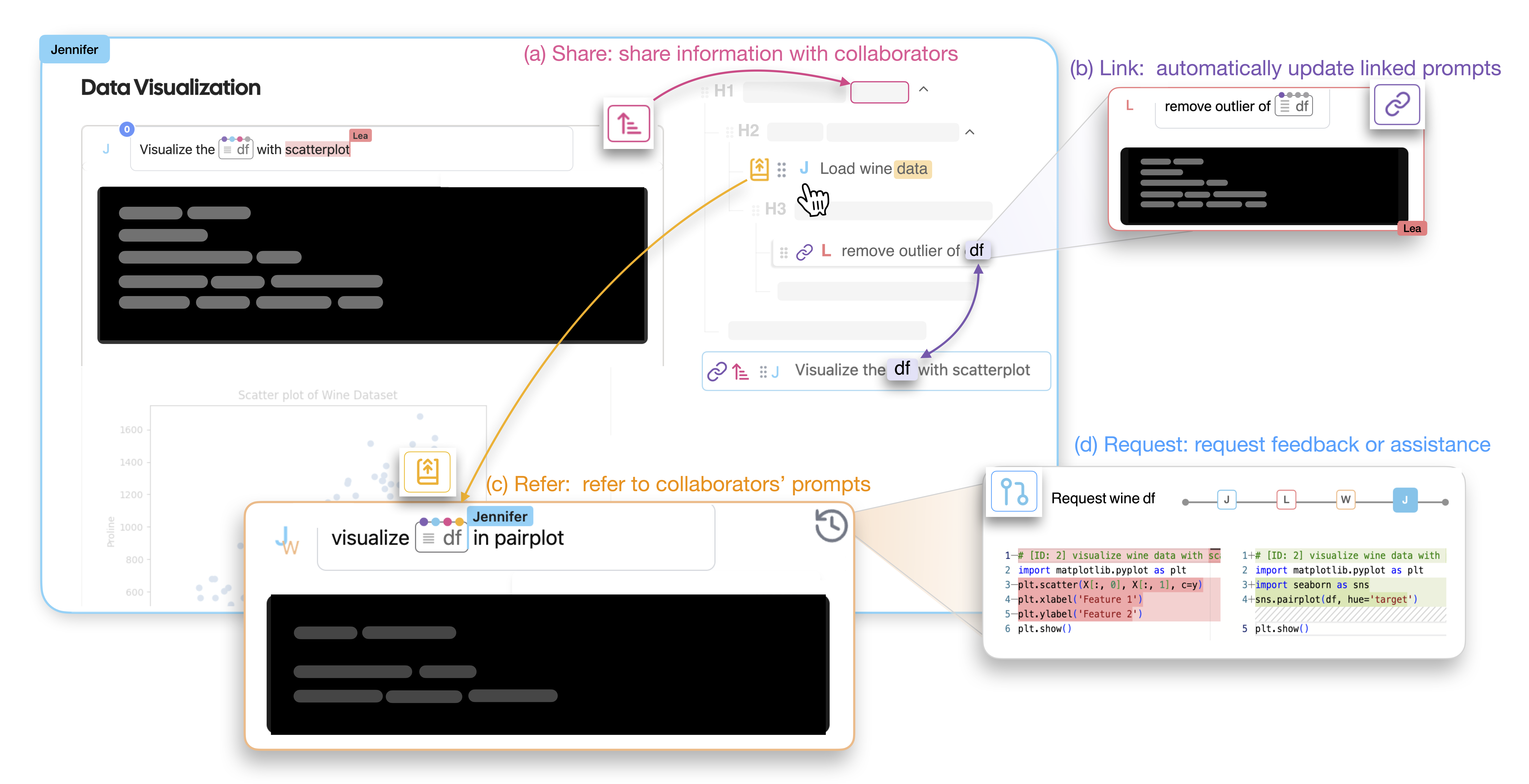}
  \caption{\textit{CoPrompt} enables programmers to conduct collaborative prompt engineering by building upon collaborators' prompts in natural language programming. It provides four mechanisms: (a) \textit{\respondlink{}} mechanism enables programmers to share information with collaborators without much effort or interrupting collaborators' work. (b) \textit{\reciprocatelink{}} mechanism automatically updates linked prompts. (c) \textit{\referlink{}} mechanism assists programmers to modify prompts regarding collaborators' prompts. (d) \textit{\requestlink{}} mechanism enables programmers to request collaborators' assistance or feedback without interrupting collaborators' progress.}
  \Description{This is a figure containing four blocks. The block on the top left corner is the main interface, which is a rich text editor. There are three programmers connected to the interface: Jennifer, William, and Lea. There is a pink arrow in this block indicating the "share" mechanism: share information with collaborators. The block on the top right corner is extended from the block on the top left corner, indicating the "link" mechanism that allows programmers to modify prompts with reference to collaborators' prompts.}
  \label{fig:overview}
\end{teaserfigure}

\maketitle

\section{Introduction}
Collaborative programming has been widely studied and supported through a range of collaborative systems \cite{Wang2019How, Wang2020Callisto, Wang2022Documentation, Fan2017Balancing, Fan2012ATCoPE, Park2018Post, Quaranta2022Eliciting, Ma2022Integrating}. 
These systems assist programmers in collaboratively writing, discussing, and debugging code across various contexts, such as data science \cite{Wang2019How, Wang2020Callisto, Wang2022Documentation, Quaranta2022Eliciting} and software development \cite{Fan2017Balancing, Fan2012ATCoPE, Park2018Post}.
While code has traditionally played a central role in collaborative programming, the emergence of large language models (LLMs) introduces an alternative approach: using natural language (NL) for programming and collaboration.
Leveraging the context offered by NL prompts allows programmers to effectively communicate with their collaborators using a more intuitive language without getting into the intricacies of low-level code~\cite{Wang2022Documentation}.
This benefit aligns with programmers' preference for understanding collaborator tasks from a high-level perspective \cite{Wang2019How}.

To obtain desired code generation results, programmers often need to conduct \textit{prompt engineering}, which involves iteratively refining prompts to guide LLMs in solving programming tasks by evaluating the generated results \cite{Reynolds2021Prompt, Liu2023Pre, Fiannaca2023Programming}.
However, prompt engineering is challenging in collaborative programming. 
To effectively collaborate with others and ensure that their engineered prompts align with the ongoing work of their collaborators, programmers need to stay informed about their collaborators' progress. This involves regularly reviewing their collaborators' code and engaging in clear communication without causing any disruptions to their collaborators. 
Nevertheless, programmers encounter difficulties when switching between their code and that of others \cite{Fan2017Balancing}, including issues such as a lack of contextual references \cite{Wang2020Callisto} and limited support for interactive sharing of intermediate results \cite{Chattopadhyay2020what, Liu2020Paths, pang2022data}.
Additionally, balancing the inclusion of contextual information in prompts can be challenging for programmers \cite{Liu2023What, sarkar2022like}. Deciding on the right amount of detail to incorporate is not always straightforward, resulting in prompts that may either lack essential information (e.g., \qt{Web Scraping}) or become overly detailed (e.g., \qt{Extract all anchor <a> tags from the parsed HTML and iterate through each}). This variation can lead to confusion among collaborators and impact the readability and reusability of prompts.
These issues increase the cognitive load of making sense of prompts and thus increase the communication cost of collaboration.

The purpose of this research is thus to explore the design of workflows that support programmers in \textit{Prompt Co-Engineering}, which involves collaboratively refining and sensemaking prompts during NL programming.
\rv{We selected data science work as a case to demonstrate the potential workflows of prompt co-engineering. Given the exploratory and explanatory nature of data science, it requires programmers to collaborate closely by sharing intermediate results~\cite{Chattopadhyay2020what, Liu2020Paths, pang2022data} and engaging in discussions with the assistance of contextual references~\cite{Wang2020Callisto}. However, our primary focus is to comprehend and facilitate the workflow of prompt co-engineering rather than addressing specific data science tasks.
} 

We conducted a formative study to gain insights into potential prompt co-engineering workflows and challenges.
Our findings revealed that programmers struggled to maintain a shared common ground, track collaborators' revision histories of prompts and code, and comprehensively understand the code solely from prompts due to their iterative nature.
They also encountered challenges in managing the procedural dependencies \cite{Liu2021Boba, Sarma2023Multiverse} when dealing with variables represented by different NL prompts, which resulted in repetitive updates.
These findings highlight the importance of supporting the prompt co-engineering workflow that encompasses comprehending collaborators' work and leveraging it to construct their own prompts.

Building on these findings, we introduce four innovative mechanisms for natural language programming aimed at reducing the effort required to build upon others' work and facilitate information sharing among collaborators: \textit{referring}, \textit{requesting}, \textit{sharing}, and \textit{linking} (\autoref{fig:overview}).
\textit{Referring} enables programmers to locate and access their collaborators' prompts by presenting user-defined tasks and prompts within a shared multi-level hierarchy view of prompts.
\textit{Requesting} and \textit{sharing} enable programmers to share information with their collaborators and solicit feedback to enhance their prompts. 
The \textit{linking} mechanism facilitates automatic updates for elements with procedural dependency, reducing the necessity for repetitive prompt modifications.

Incorporating these mechanisms, we designed \sys{}, a prototype system that assists the workflow of \textit{Prompt Co-Engineering}. \sys{} supports programmers in making sense of collaborators' work with multi-level hierarchical interactions and contextual prompt information, as well as leveraging collaborators' work and sharing information.
To evaluate the usefulness of \sys{} in assisting prompt co-engineering workflow during collaborative NL programming, we conducted a 2-part user study involving 12 experienced programmers familiar with LLM-based code assistants and collaborative programming.
\rv{
Participants were asked to complete a real-time collaborative programming task in pairs, working together simultaneously. They were then tasked with following up on the work of other participant pairs to simulate asynchronous collaboration. This involved comprehending and modifying existing work by others and independently addressing tasks without online collaboration.
}
The results showed that \sys{} effectively supported programmers in understanding their collaborators' prompts and facilitated communication among them to build upon each other's work. 
In summary, this research makes the following contributions:
\begin{itemize}
    \item A formative study that uncovered the workflow and challenges of collaborative NL programming.
    \item \sys{}, a prototype system that supports programmers' prompt co-engineering workflow during collaborative NL programming by comprehending, referring, requesting, sharing, and linking with collaborators' prompts.
    \item A user study that provided insights into the usability and usefulness of \sys{} and design implications for future systems assisting prompt engineering in collaborative NL programming contexts.
  \end{itemize}

\section{Related Work}
As our research aims to address the challenges of collaborative prompt engineering in collaborative NL programming, we review prior work on natural language programming and LLMs, as well as collaborative programming groupware.

\subsection{Natural Language Programming and Prompt Engineering}
Natural Language (NL) Programming is the process of using NL to express programming ideas for desired output \cite{Miller1981Natural, mihalcea2006nlp}. Prior work provided insights on how people express computer-like procedures ``naturally'' and on what features programming languages should include to be more ``natural-like'' \cite{Miller1981Natural}. 
With the development of natural language processing (NLP) \cite{chowdhary2020natural}, it has become feasible to use NL to conduct more programming tasks, as it allows more free-form NL utterances to be translated into program code \cite{Liu2005Programmatic}. This advancement increased the accessibility of programming to non-expert users \cite{mihalcea2006nlp} and end-user programmers \cite{Ko2011The} who lack training in computing.

Recently, the advances in generative AI \cite{Muller2023GenAICHI, wang2023reelframe1}, especially LLMs \cite{chen2021evaluating}, fostered the capability of generating code from NL prompts, by allowing a wider space of utterances to be transformed into satisfying code snippets. This advance significantly enhanced the performance of AI-driven code assistants \cite{Github2022Copilot} and thus improved the satisfaction and accessibility of programming with NL prompts \cite{Jiang2022PromptMaker, sarkar2022like}. LLM-powered code assistants allow programmers to write at different levels of abstraction when developing code, which provides a greater degree of freedom \cite{sarkar2022like, Henley2014The}. 
\rv{Prior work has investigated the design space of AI-powered code assistants for computational notebooks ~\cite{Mcnutt2023On} and other popular code editors~\cite{Vaithilingam2023Towards}.
}
However, the multiple levels of abstractions \cite{green1996usability} in the NL prompts also resulted in the abstraction matching problem when using NL for programming, where programmers find it difficult to select an utterance that will translate into the desired system action \cite{yen2023coladder, sarkar2022like, Pearce2023Examining, Luger2016Like, Jiang2022Discovering}. 
A case study investigating the NL prompting process of prototyping also highlighted the difficulty of evaluating whether a prompt is improving \cite{Jiang2022PromptMaker}. This issue is rooted in the challenges of translating user instructions into executable computer tasks \cite{hutchins1985direct}.

To mitigate the abstraction matching issue, prior work has investigated ways of prompt engineering, which is the process of engineering an NL prompt to make it more effective in generating desired results \cite{brown2020language, Reynolds2021Prompt}. Liu et al. proposed design guidelines for prompt engineering for text-to-image generative models \cite{Liu2022Design}.
Common practices in prompt engineering include appending information like explanations \cite{lampinen2022language}, demonstrations \cite{cypher1993watch, lieberman2001your}, table schema \cite{trummer2022codexdb}, and relevant examples (few-shot prompts) \cite{Liu2023Pre, Jain2022Jigsaw}. Although few-shot prompts have become a popular strategy, they may still behave worse than zero-shot prompts sometimes \cite{Reynolds2021Prompt}. To enhance the effectiveness of the prompts, programmers can further specify tasks by constructing the signifier, memetic proxy, and specifying truth-seeking patterns \cite{Reynolds2021Prompt}. 

Other prompt engineering methods include combining specific task information with general intentions (meta-prompts) \cite{Reynolds2021Prompt}, generating mutations of the prompt \cite{Li2023CCTEST}, eliciting feedback with small data \cite{strobelt2023interactive}, summarizing complicated prompts \cite{kuznia2022more}, defining prompt grammar \cite{Fiannaca2023Programming}, \rv{uncertainty highlighting~\cite{vasconcelos2023generation}} and introducing a new programming language \cite{Beurer2023Prompting, huang2023anpl}. 
Prior research into natural language interfaces suggests the benefit of managing expectations and gradually revealing the capabilities of the system through user interaction and intervention \cite{ross2023programmer, vasconcelos2023generation}. There are also practices of breaking down tasks \cite{Ritschel2022Can} and dividing complex tasks into chained series of sub-tasks \cite{Wu2022AI}. By breaking down complex problems into sub-tasks, the gap in abstraction is reduced, enabling successful guidance of the model to generate code that matches the programmer's intents \cite{bodker2015third}. 

However, prompt engineering in collaborative programming, which involves understanding and utilizing collaborators' work, remains unexplored. This work aims to investigate the challenges and benefits of NL prompts with varying levels of abstraction in collaborative contexts.

\subsection{Collaborative Programming}
Extensive research in HCI and CSCW has investigated challenges and system designs to assist collaborative programming. 
Synchronization is a challenging yet significant part of the collaboration, as programmers need to synchronize with their collaborators in various artifacts like data frames, variables, and archives \cite{Wang2019How}. It is challenging as the artifacts in programming involve procedural dependencies \cite{Liu2021Boba, Sarma2023Multiverse}: if one part of the code changes, all related code snippets must be updated to prevent conflict and errors. In addition, programmers often encounter difficulties when switching between their code and that of others \cite{Fan2017Balancing}.
\rv{The problems of context switching and knowledge sharing are common, especially in the data science domain where programmers need to frequently share intermediate results \cite{Chattopadhyay2020what, Liu2020Paths, pang2022data} and discuss with the assistance of contextual references \cite{Wang2020Callisto}.}

Establishing group awareness~\cite{dourish1992awareness} can reduce communication costs and thus improve collaboration efficiency \cite{Wang2022Improving}. It involves understanding the activities of others, information sharing, and knowledge of group and individual contexts \cite{gutwin1998effects, lauwers1990collaboration, Gutwin2002Descriptive, herbsleb1995object, Muller2019How}. 
\rv{This is particularly important yet challenging in the domain of data science due to the diversity of artifacts and individuals involved in data science work~\cite{Crisan2021Passing}.}
To facilitate comprehending the complex dependencies and relationships among collaborators' work, Albireo displays the relationships between the cells of a computational notebook using a dynamic graph structure \cite{Wenskovitch2019Albireo}. 
Documentation plays an important role in maintaining shared understanding and group awareness \cite{de2005A, kajko2005survey, Maalej2013Patterns, Roehm2012How, shi2011empirical}. To document the development progress, programmers write comments to make the code easier for both themselves and others to understand \cite{Padioleau2009Listening}. 
Comments are also essential for sharing intermediate results \rv{in data science work}~\cite{Pang2022How}.
However, writing comments is tedious which makes many programmers not bother to write comments in time. The lack of detailed explanations and intention-revealing comments causes trouble for others understanding their work \cite{geiger2018types}. To make the commenting process easier, Wang et al. built Themisto, which leveraged AI to provide AI-assisted comments based on deep learning, query, and prompt \cite{Wang2022Documentation}. Their user study suggested that the collaboration between data scientists and Themisto significantly reduced task completion time and resulted in satisfaction.

Reusing collaborators' work is also challenging in collaborative programming \cite{Liu2021To, sarkar2022like, Wang2020From}. To facilitate referring to collaborators' work, chat.codes enabled programmers to link code with messages in the chatroom \cite{Oney2018Creating}. In addition, Codeon provides on-demand remote collaboration assistance by automatically capturing the relevant code context and allows remote helpers to respond with high-level descriptions, code snippets, and NL explanations.\cite{Chen2017Codeon}. 
Communication is essential for maintaining shared understanding and group awareness in collaborative work \cite{gutwin1998effects}. %
While it could be time-consuming in software developing collaboration \cite{herbsleb1995object}, prior work has investigated ways of reducing collaborators' communication costs through documentation~\cite{geiger2018types}, comments~\cite{Padioleau2009Listening}, visualizations, and version control systems \cite{ Chen2017Codeon}. 
\rv{Considering the heavy dependencies among the artifacts involved in data science work, code-gathering tools highlight dependencies used to compute results to assist programmers in understanding, reusing, and rewriting in cluttered notebooks~\cite{Head2019Managing}.}

However, these collaboration systems have not considered NL programming, where the challenges of comprehending and leveraging collaborators' work are different. \sys{} aims to investigate the challenges and potential solutions for the challenges in collaborative NL programming, especially prompt co-engineering.

\section{Formative Study}
We conducted a formative study to understand the challenges faced by programmers and their needs in the workflow of prompt co-engineering. Specifically, we focused on how programmers comprehend and build upon their collaborators' work to iteratively refine their prompts for generating code that matches their intents. 

\subsection{Participants and procedure}
Five pairs of experienced programmers familiar with LLM-based code assistants were recruited. Participants reported 6–24 months of experience with an AI code assistant and 3–5 years of experience using computational notebooks. All participants were 20–35 years old and had at least bachelor’s degrees in a CS-related field. 

We asked participants to work remotely in pairs on an exploratory data programming task using a shared Jupyter notebook in the VSCode editor \cite{Microsoft_2021a} embedded with the GitHub CoPilot plugin.
\rv{The real-time sharing functions are provided by the \textit{Live Share} plugin~\cite{Microsoft_2021}, which synchronizes edits between users and allows collaborators to see each other’s cursors. The data programming task \cite{houseprices} is a popular data science task on the Kaggle platform which requires participants to use advanced regression techniques to conduct a prediction. It involves common data science operations such as data cleaning, feature transformation, and correlation analysis.}

Participants were asked to join a Zoom meeting first to discuss their task distribution and collaboration workflow. 
Then, they started working on their own tasks using NL prompts, and they were allowed to communicate via audio in the meantime. 
\rv{To observe the natural prompt co-engineering workflow, participants were explicitly required to modify the NL prompts instead of directly tweaking the code.
While the study session was conducted in real-time, the nature of the tasks did not require synchronous collaboration, i.e., participants had the flexibility to divide the tasks into sub-tasks and work on them asynchronously. We did not explicitly require participants to complete the tasks synchronously or asynchronously. 
}
\rv{The collaborative programming session lasted about 90 minutes, after which participants were asked to attend 30-minute follow-up interviews. 
We instructed participants to try their best to complete the tasks in high quality and efficiency.
While there was no external incentive for the task performance, all participants successfully completed the tasks.} 
All participants received compensation according to local standards.

\subsection{Data Analysis}
\rv{
The studies were logged using VSCode extensions and the process was video-recorded and transcribed. Two co-authors conducted an inductive thematic analysis~\cite{braun2012thematic} involving cross-referencing timestamped data of prompt modifications from system logs with video recordings and interview transcripts, identifying the events that transpired before prompt engineering. 
Specifically, the two co-authors read through the transcripts first to familiarize themselves with the data and then performed the open coding process independently. 
Then, all co-authors discussed and updated the code book during the weekly project meeting for two weeks.
Finally, we categorized and analyzed a total of 229 instances out of 392 recorded interactions between participants, which fall into 3 stages in the prompt co-engineering workflow: comprehension, pre-modification interactions, and prompt modification}. We excluded 163 actions due to a lack of clear context or relevance to the specific collaborative communication events that preceded prompt modifications.

\begin{table*}[ht]
\centering
\footnotesize
\begin{tabular}{m{2.7cm}m{8.3cm}m{4cm}r}
\specialrule{.2em}{0em}{.2em} %
\textbf{Action} & \textbf{Description} & \textbf{Example} & \textit{n} \\
\specialrule{.15em}{.25em}{.15em} %
Sync Up & Participants communicated to align their efforts and ensure consistency in their tasks. These interactions helped prevent redundancy and maintain cohesiveness in their prompt engineering process. & ``I am encoding the whole dataset." & 49 \\
\hline
Reactive Communication (Request \& Response) & These interactions aim to seek assistance, feedback, or validation regarding reactively handling specific aspects of the prompt or generated code. & ``Can someone help me determine the function for encoding?" & 61 \\ 
\hline
Clarifying & Participants seek answers to queries about their collaborators' work, including seeking explanations, and verifying the correctness of specific code segments. & ``Did you drop the column of xxx?" & 57 \\
\hline
Reference \& Reuse & Participants occasionally referred to and reused (e.g., copy-paste) components from their collaborators' work, utilizing these references to inform their own prompt modifications. These actions fostered a sense of collaboration and knowledge exchange. & ``I used the function you wrote in my block. Any concerns?" & 33 \\
\hline
Proactive Communication & Proactive communication involves participants sharing insights, updates, or information related to their prompts or coding tasks. These exchanges often contributed to a deeper understanding of the prompt's context. & ``Here's an update on the changes I made to the prompt..." & 29 \\
\specialrule{.15em}{.15em}{.15em}
\end{tabular}
\caption{Five types of actions of the pre-modification interactions.}
\Description{This is a table including four columns: Actions, Description, Example, and "n", there are five rows in the table, indicating five types of actions.}
\label{table:pre-modification}
\end{table*}

\subsection{General Workflows in Prompt Co-Engineering}
In the following paragraphs, we present our findings of the workflows that participants adopted in prompt co-engineering and the challenges \textbf{(C)} that they encountered.

In the initial stages, participants convened online to gain an understanding of the data. Subsequently, high-level task distribution was discussed and noted down at the beginning of a collaborative computational notebook. We noticed that 4 pairs structured the task by hierarchically numbering lists or bullet points to externalize the task structure in their minds.
Upon settling on a preliminary task distribution, participants started writing prompts independently to accomplish their own distributed tasks. Throughout the process, they maintained communication via Zoom to discuss ongoing and potential code implementations. We observed that all participants started with a high-level description of the task (e.g., data preprocessing, encoding) with the methods involved (e.g., low-pass filtering, linear regression), deliberately omitting details (e.g., variable names and parameters). Participants mentioned the reason is that they have to \pqt{wait until her [collaborator's] work is complete}{P4} to continue adding more details to the prompt. Therefore, participants do not verify the generated code in detail at first because they know that \pqt{it will eventually be changed later}{P2}. To further refine their prompts, participants primarily go through three stages:

\paragraph{\textbf{Stage 1 --- Comprehension}:} Regularly checking in on collaborators' work became a common practice.
The goal was either to reuse collaborators' work or to assess their progress and determine the next steps. However, significant time was spent \pqt{scrolling up and down to identify changes}{P9}. 
Participants also encountered difficulties in comprehending their collaborators' code based solely on the prompt, often describing the prompts as \qt{unorganized} and \qt{vague}. Additionally, P7 highlighted another issue where the generated code \qt{sometimes not aligned with the prompt,} further complicating the comprehension process. Lastly, participants faced challenges in tracking their collaborators' 
revision history of prompt and code, which is essential for understanding \pqt{the reasoning behind code changes}{P1} (\textbf{C1}).

\paragraph{\textbf{Stage 2 --- Pre-Modification Interactions}:} Programmers often adapt their prompts based on their own experience and the current work, 
which can be challenging to document comprehensively. We thus focus on the explicit collaborative strategies employed prior to the start of prompt engineering. Based on the thematic analysis results, these pre-modification interactions consisted of a series of actions (Table \ref{table:pre-modification}).

\paragraph{Syncing up with collaborators.} 
Many participants found that their initial task distribution was not detailed enough, which caused redundant effort and inappropriate workflow between collaborators: \pqt{We encoded the data at the same time}{P3}.
Participants also reported the tedious process of updating prompts due to procedural dependency \cite{Liu2020Paths}, in which a downstream prompt only works if a particular upstream prompt works normally. 
Due to the ever-changing nature of data programming work, programmers often need to monitor their collaborators' changes and update their prompts to align with their collaborators' process, otherwise, they may \pqt{encounter error messages due to collaborators' modifying the data frame halfway through the process}{P2} (\textbf{C2}).

\paragraph{Requesting for collaborators' feedback and assistance.}
All participants in charge of visualizing correlation (P1, 4, 5, 8, 9) left comments asking for feedback, as it is an essential step for data analysis.
We also observed that some participants (P3, 5, 8, 9) requested help from their collaborators.
For instance, P3 asked his collaborator to handle a sub-task that he failed to complete.
There are also cases where collaborators need to work closely and go through a trial-and-error process together: \pqt{I asked my collaborator to pay attention to the outliers every time the way of feature transformation is changed}{P4}. However, most participants (N=7) indicated that they desire a non-interruptive method to send their collaborators requests, instead of speaking up, which is too interruptive for them to use frequently (\textbf{C3}).

\paragraph{Referring to collaborators' processes and prompts.}
All participants checked their collaborators' processes and referred to their prompts to improve their own for better generation results across the whole notebook. To leverage others' prompts, participants first locate and read the target prompt to make sense of it. Then, they copy portions of the prompt relevant to their task and integrate them into their own prompts to provide contextual information.
The redundant copy-pasting and modifying can be time-consuming (\textbf{C4}), as programmers may trial-and-error to determine the appropriate modifications of the prompts \textit{(P1: ``I reused my collaborator's prompt, which did not work as I imagined. After analyzing its context, I realized that I had to copy a prompt several blocks above that'')}.

\paragraph{Proactive communication for sharing intermediate results and relevant information}
Many participants have shared intermediate results with their collaborators that they believe would be useful. They performed three types of sharing strategies: (1) leaving comments under the block that their collaborators were working on to attract their attention - P3; (2) leaving comments before the block of the shared information and pinning their collaborators using an ``@'' - P5, 6; and (3) ask their collaborators to check their current highlights block for reference - P1, 2, 9. The first strategy requires the comment receiver to locate the shared information, while the second strategy may influence the collaboration efficiency. Though many participants communicated directly through Zoom, it \pqt{disturbed my own progress a bit}{P10}. These strategies are either \pqt{inefficient}{P9} or \pqt{disruptive}{P7} (\textbf{C4}).

\paragraph{\textbf{Stage 3 --- Prompt Modification \& Merge Conflicts}}
The third stage centers on modifying (i.e., engineering) the prompt. 
In this stage, participants refine their prompts by copying and pasting utterances or code snippets from collaborators' prompts to clarify details about the variable name, resource, methods, and detailed considerations.
During this phase, participants may encounter merge conflicts or issues that need communication for resolution. Participants also expressed a desire to access previous versions of the code, as this helps them \pqt{recall who made specific changes to the prompt}{P4} and the \pqt{reasons behind those alterations}{P7}. \rv{Some participants manually tweaked the code instead of modifying the NL prompts when they could not achieve their desired results within a few attempts: \pqt{find it more efficient to directly change code after several failed trials}{P1}}.

In summary, we identified the following user challenges in the formative study:
\begin{itemize}
    \item \textbf{C1}: Effort of maintaining group awareness and shared understanding to enhance collaboration effectiveness.
    \item \textbf{C2}: Repetitive effort of syncing with collaborators' work.
    \item \textbf{C3}: Inconvenient and disruptive ways of requesting collaborators’ feedback and assistance.
    \item \textbf{C4}: Repetitive copy-pasting effort for leveraging others’ work and disruptive information sharing.
\end{itemize}

\section{Design Considerations}

Based on the findings from the formative study, we formulated four Design Considerations (\textbf{Ds}). to support prompt co-engineering in collaborative NL programming.

\textbf{D1: Supporting sense-making of collaborators’ progress and prompts.}
Programmers encountered challenges locating collaborators' work in a shared notebook that lacked a clear outline of NL prompts and code (C1).
To support programmers' locating and sense-making of collaborators' progress, it is important to implement a clearer structure to show the notebook overview~\cite{Fiannaca2023Programming}. This structure should include multiple levels of hierarchy (e.g., tasks, sub-tasks, and prompts) to help programmers identify changes easily. The design should also incorporate assistive features to help programmers understand the code from prompts that might be too vague.
In addition, a history view should be provided for programmers to track global activities (e.g., collaborators' works) and local changes (e.g., prompts variations).

\textbf{D2: Automatic synchronization to reduce repetitive updates.}
In the exploratory and iterative programming process, the prompts and codes might be updated several times throughout the whole process based on the collaborators' changes (C2).
Automatic synchronization of variables and code snippets with procedural dependency should be provided to reduce programmers' cognitive load and enhance their collaboration efficiency. Programmers should be able to easily notice changes made by collaborators and the automatic updates applied to their own work.

\textbf{D3: Supporting requests for feedback and assistance.}
Current ways of requesting collaborators’ feedback and assistance are inconvenient and disruptive (C3). An efficient way 
of requesting feedback should be provided besides communicating through chat and voice.
In addition, considering the situation that the collaborator has not finished the required prompt, programmers should be equipped with methods to request knowledge from others and refer to it later.

\textbf{D4: Reduce effort for sharing knowledge and incorporating others’ work.}
Current ways of referring to collaborators' prompts and reuse are tedious and time-consuming (C4). Programmers need to copy, paste, and modify, which takes a lot of time unnecessarily. A more effortless way of referring to collaborators' prompts should be provided. Additionally, the design should enable proactive sharing of intermediate results with collaborators, promoting sharing without concerns about disrupting their workflow.

\section{Envisioned Scenario}
\begin{figure*}[h]
    \centering
    \includegraphics[width=0.9\linewidth]{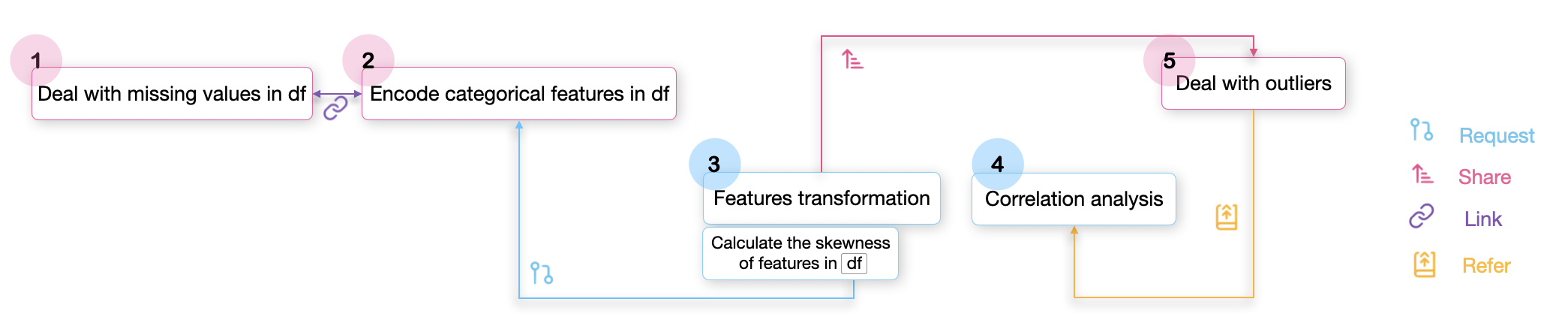}
    \caption{Envisioned Scenario of collaborative NL programming using \sys{}, including tasks from 1 to 5 (pink boxes indicate Alice's tasks, and blue boxes indicate Bob's tasks). Four colors of arrows indicate four types of mechanisms.}
    \Description{This figure contains five blocks, labeled 1 to 5. Block 1 writes "Deal with missing values", Block 2 writes "Encode categorical features in df", Block 3 writes "Features transformation", Block 4 writes "Correlation analysis" and Block 5 writes "Deal with outliers".  Blocks 1, 2, and 5 have pink borders, which indicates that these blocks are Alice's tasks, Blocks 3 and 4 have blue borders, which indicates that these blocks are Bob's tasks.}
    \label{fig:scenario}
\end{figure*}

Here, we present a motivating scenario that illustrates the workflow of using \sys{} for prompt co-engineering in NL programming.
For simplicity, we describe our scenario using two collaborators, Alice and Bob.

Alice and Bob are remotely collaborating on a data analysis task requiring them to predict house prices. 
To improve collaboration efficiency, they divide the tasks into smaller segments, allowing each to tackle different tasks separately.
They decide that Alice would handle missing values, outliers and categorical features. 
Meanwhile, Bob is tasked with feature transformation and correlation analysis (\autoref{fig:scenario}). 
To track each other's progress and offer/request help for specific tasks, they outline all sub-tasks in \sys{}'s rich text editor (\autoref{fig:UI} a), which is synchronously displayed in the multi-hierarchical wiki (\autoref{fig:UI} b).
The wiki's foldable task items provide them with a clear overview of tasks and the collaboration process.
They then begin to work on writing prompts independently after listing all the required tasks.

\begin{figure*}[h]
    \centering
    \includegraphics[width=\linewidth]{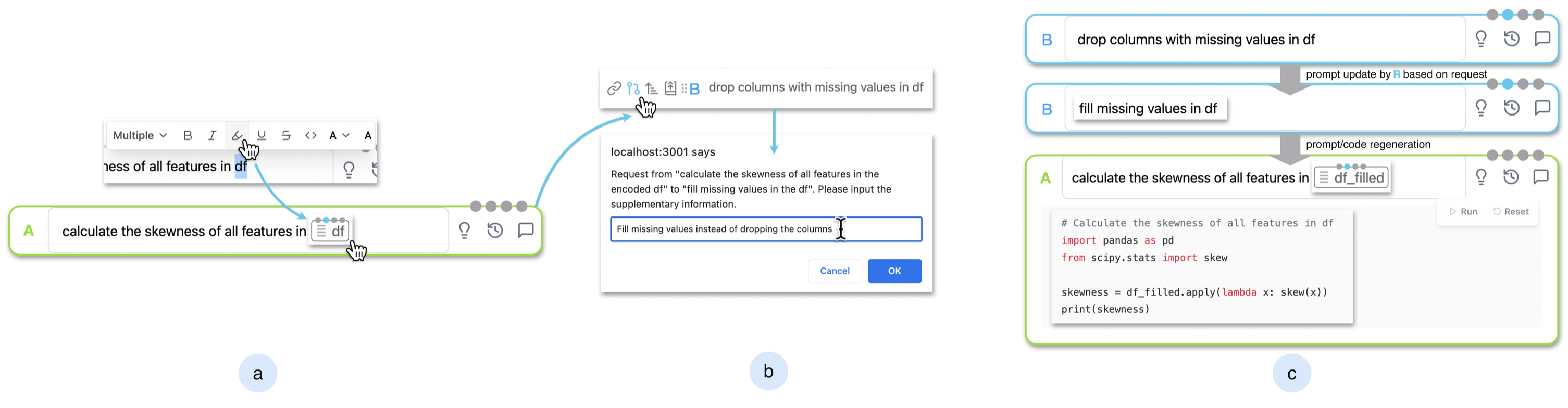}
    \caption{Request Workflow: (a) select a source node from the editor; (b) select a target node that needs the collaborator to finish and share the result from wiki and click the \requestlink{} icon, fill in a descriptive message for the request; (c) when B updates how they deal with missing values, A's block will regenerate prompt and code accordingly.}
    \Description{There are three parts (a, b, c) of this figure for the workflow of request (from left to right): (a) select and highlight prompt utterances that should be modified from the editor as a node; (b) select a target prompt of the task that needs the collaborator to finish and share the result from wiki and click the request icon, fill in a descriptive message for the request; (c) when B updates how they deal with missing values, A's block will regenerate prompt and code accordingly.}
    \label{fig:requestlink}
\end{figure*}

Although Alice plans to finish encoding before Bob begins feature transformation, her progress is delayed due to technical issues.
Due to time constraints, Bob cannot wait for Alice to finish encoding.
With \sys{}, Bob creates a \includegraphics[width=0.025\linewidth]{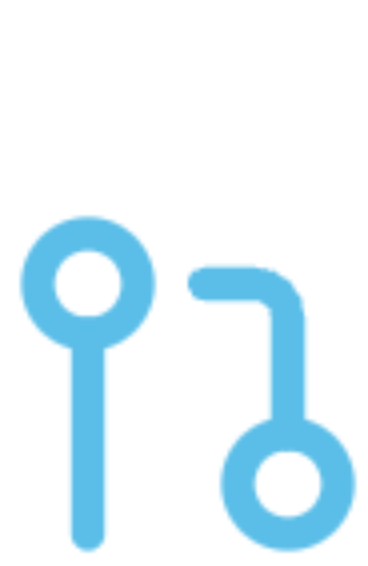} \textcolor{cyan}{\textbf{\requestlink{}}} (\autoref{fig:requestlink}) from the \includegraphics[width=0.043\linewidth]{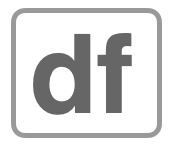} in the transformation prompt to Alice's encoding prompt, indicating that his feature transformation steps (e.g., calculating the skewness) require Alice's encoded result.
Once \sys{} detects (through semantic analysis) that Alice has completed the encoding, it automatically updates Bob's prompt to leverage encoded data to generate code. 
As a result, Bob no longer needs to manually modify his prompt whenever Alice updates her encoding, saving his time and allowing him to focus more on task completion rather than repetitive code upkeep.

After completing the feature transformation, Bob checks the progress from the wiki (\autoref{fig:UI}) and notices that Alice needs the transformed data for later outlier handling. In case Alice needs to spend much time verifying the data to be used, Bob decides to proactively share it with Alice.
To share this transformed data, Bob utilizes the \includegraphics[width=0.025\linewidth]{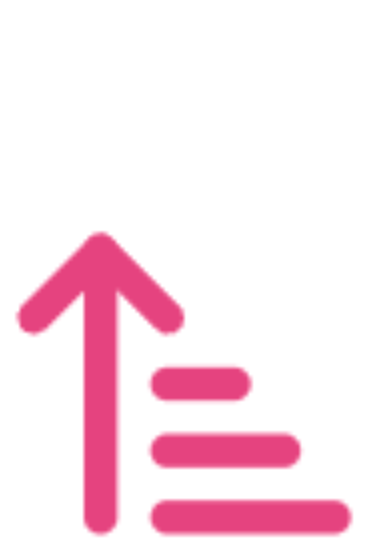} \textcolor{magenta}{\textbf{\respondlink{}}} mechanism (\autoref{fig:passlink}) by highlighting his data frame and clicking on the share icon next to Alice's task in the wiki. 
Alice receives a pop-up, allowing her to accept Bob's data without manually locating the data frame for her subsequent workflow.
Once accepted, \sys{} automatically regenerates Alice's prompts based on Bob's shared dataframe, eliminating the need for Alice to input supplementary information like variable names. 
This reduces the risk of human errors such as incorrect variable references.

\begin{figure*}[h]
    \centering
    \includegraphics[width=\linewidth]{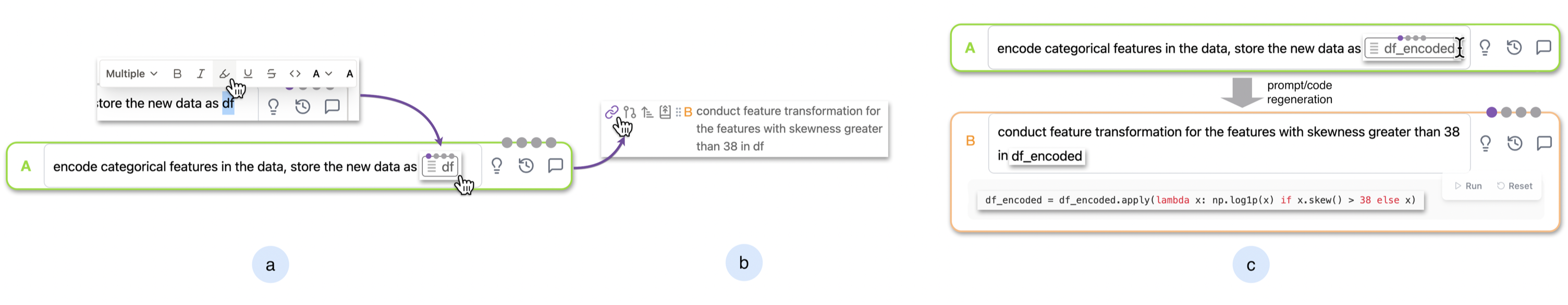}
    \caption{Link Workflow: (a) select a source node from the editor; (b) select a target node to be synced with the source node from the wiki and click the \reciprocatelink{} icon; (c) when A updates the name of the df, B's block will regenerate prompt/code accordingly.}
    \Description{There are three parts (a, b, c) of this figure for the workflow of link (from left to right): (a) select and highlight the element to be synced as a node from the editor; (b) select target prompt that is to be synced with the highlighted element from wiki and click the link icon; (c) when A updates the name of the df, B's block will regenerate prompt/code accordingly.”}
    \label{fig:directlink}
\end{figure*}

After handling missing values and encoding, Alice anticipates that there may be future adjustments to the missing value handling based on her past experience, which may also require updates to her encoding methods.
To avoid repetitive updating, Alice creates a \includegraphics[width=0.025\linewidth]{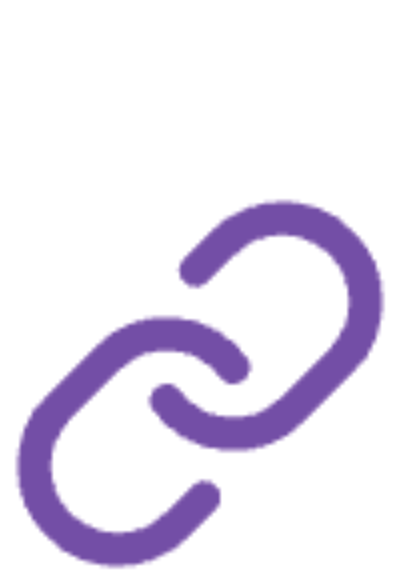} \textcolor{violet}{\textbf{\reciprocatelink{}}} (\autoref{fig:directlink}) connecting the 
\includegraphics[width=0.043\linewidth]{figures/df.png} in the missing value handling prompt and the \includegraphics[width=0.043\linewidth]{figures/df.png} in the encoding prompt. 
With this link, the encoding prompt gets automatically updated whenever the prompt for missing value handling changes (e.g., when \includegraphics[width=0.043\linewidth]{figures/df.png}'s variable name or the method for handling missing values changes). 
This way, Alice avoids the need to repeatedly update the encoding prompt. 
\rv{If Alice no longer wants the two nodes to be automatically synced, she can unlink the nodes by de-highlighting the link icon.}

When handling outliers, Alice needs to refer to the results of Bob's correlation analysis. 
However, she finds it challenging to navigate through Bob's prompts, which contain long execution examples and demonstrative code.
To help her better understand Bob's prompts, Alice expands the explanation view (\autoref{fig:UI}d) to see the highlighted prompt and annotated relationships between the NL prompts and code snippets.
From the explanation view, Alice understands Bob's considerations for correlation analysis and appropriate criteria for determining outliers.
Then, Alice begins writing her prompts to address the outliers. 
To ensure no information (e.g., criteria for determining outliers) is overlooked, Alice wants to instruct the LLM to determine the outlier handling method based on the results of the correlation analysis.
Alice employs the \includegraphics[width=0.025\linewidth]{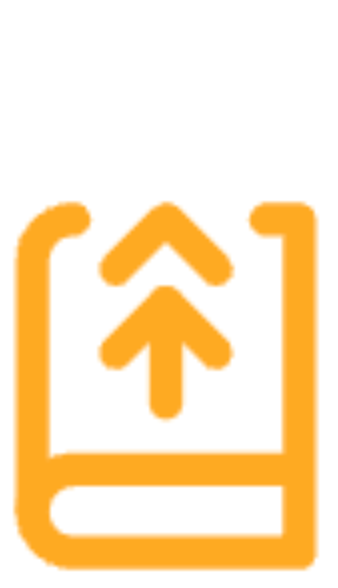} \textcolor{orange}{\textbf{\referlink{}}} mechanism (\autoref{fig:referlink}), creating a node that links the outlier handling prompt to the correlation analysis prompt. 
As a result, \sys{} automatically updates Alice's prompts with appropriate methods, no longer requiring Alice to modify her prompts by copy-pasting and typing.

\rv{In asynchronous collaboration settings, the \requestlink{}, \reciprocatelink{} and \referlink{} mechanisms work similarly as those in synchronous settings because these mechanisms do not require the collaborators to respond in real-time. 
However, there are some differences for the \respondlink{} mechanism.
Specifically, if Alice is offline when Bob is sharing artifacts, \sys{} retains the information and systematically presents each shared artifact when Alice reconnects online. 
Any modifications made are highlighted and are readily accessible through the message panel to facilitate convenient inspection. 
}
\section{Designing COPROMPT}
\begin{figure*}[h]
    \centering
    \includegraphics[width=0.95\linewidth]{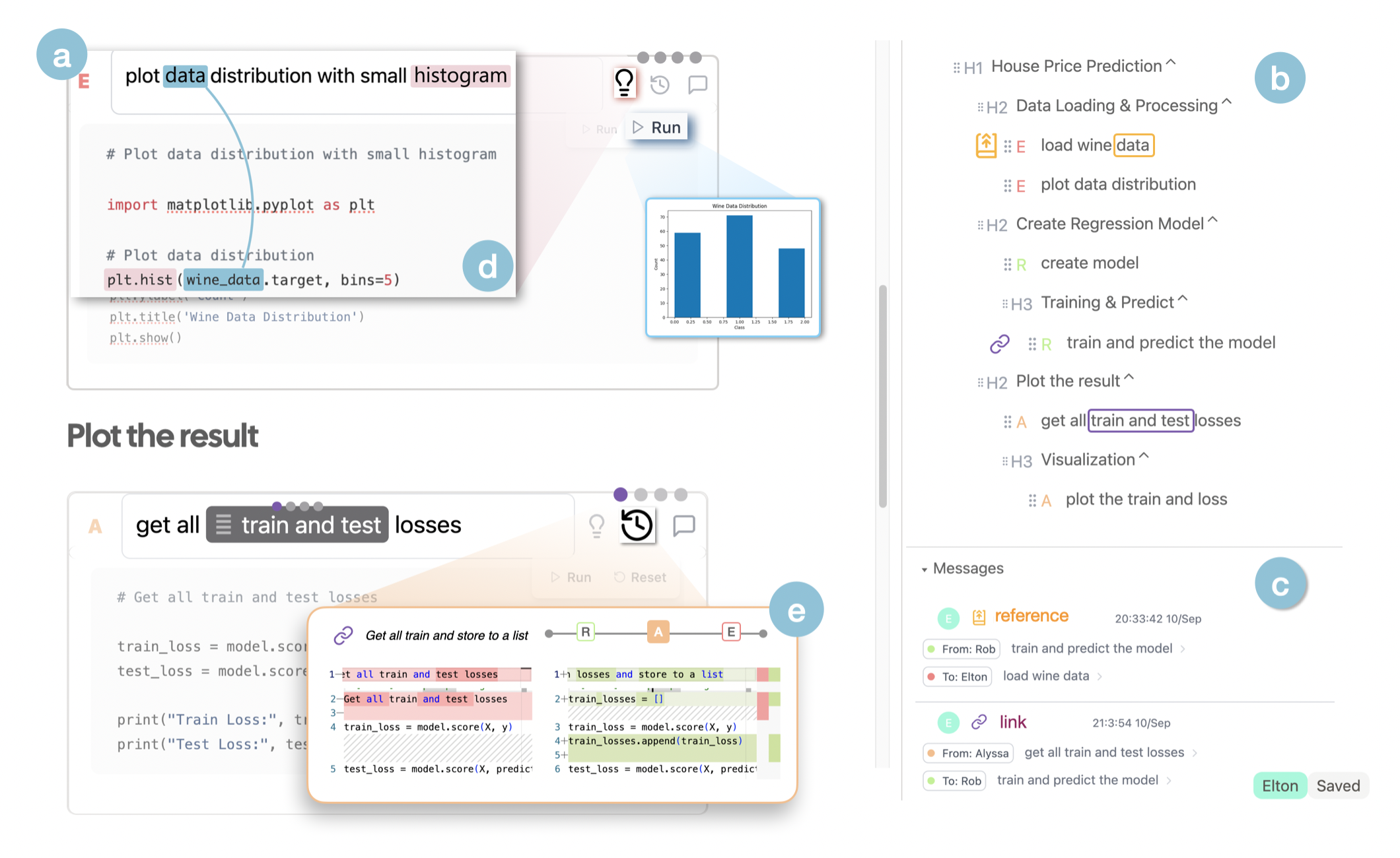}
    \caption{The \sys{} user interface includes (a) a block-based rich text editor for NL inputs, which consists of prompt blocks and execution code blocks. (b) \textbf{Prompt Wiki}: a multi-hierarchy wiki displaying tasks and prompts, (c) a \textbf{message panel} providing a comprehensive log of actions, (d) an \textbf{explanation view} displaying an explanation for prompts and code-prompt relationship, and (e) a \textbf{history view} for version control.}
    \Description{The CoPrompt user interface includes (a) a block-based rich text editor for NL inputs and operation for four mechanisms: refer, request, share and link; (b) a multi-hierarchy wiki displaying tasks and prompts; (c) a message panel providing a comprehensive log of actions; (d) an explanation view displaying explanation for prompts and code-prompt relationship; and (e) a history view for version control.}
    \label{fig:UI}
\end{figure*}

Based on our design considerations, we developed a prototype, \sys{}, to support programmers in their prompt engineering workflow during collaborative NL programming.
Specifically, \sys{} supports sense-making of the collaboration process and prompt co-engineering. \sys{} interface consists of five components (\autoref{fig:UI}): 
block-based rich text editor, prompt wiki, message panel, explanation view and history view.
\sys{} is also designed with a set of real-time collaborative features, such as real-time displays of collaborators' cursor location and their text selections.

\subsection{Sense-Making and Tracking of the Collaboration Process}
\sys{}introduces two custom block types: the \textbf{Prompt Block}, for creating prompts that generate code segments, and the \textbf{Execution Code Block}, which contains generated code that can be compiled and executed to check the interim results.
As changes are made in the text editor, the wiki view automatically updates to display the editor's structure with a tree-based representation. 
This structure provides a clear overview of the editor, allowing programmers to visualize multiple levels of hierarchy, from headings to prompts, down to individual \textit{nodes} (i.e., phrases within the prompts). 
The wiki allows intuitive navigation of its content by enabling programmers to click and fold each task item, collapsing unrelated tasks and concentrating on those of interest.
Furthermore, it provides a clear overview of the hierarchical relationships within the project and keeps programmers updated about modifications made by their collaborators (\textbf{D1}).

\paragraph{\textbf{Message Panel}}
The message panel (\autoref{fig:UI} c) offers a comprehensive log of actions, allowing programmers to track their collaborative activities. 
These actions are tied to specific elements within the editor, such as prompts and nodes, which can be easily accessed by selecting them directly from the messages.
Additionally, programmers can quickly identify essential information related to each action, including its type, creator, and timestamp.
Messages are displayed in chronological order, with unprocessed actions prioritized at the top and highlighted by a small dot. 
Once addressed, \sys{} automatically updates the panel by removing the highlights to indicate that the action was resolved.

\paragraph{\textbf{History View}}
The history view (\autoref{fig:UI} e) enhances version control and tracking of changes for prompts and their associated code.
Programmers can review previous versions of selected prompts, organized chronologically. 
This view tracks the evolution of prompts, providing details such as actions that led to changes, individuals responsible for modifications, and alterations made to the prompts. 
Similar to Git version control, \sys{} offers a diff-view that allows programmers to identify differences between current and previous versions of both prompts and code segments.

\subsection{Supporting Programmers' Prompt Co-engineering}
To support the prompt co-engineering workflow, \sys{} provides functions commonly used in collaboration interfaces (e.g., annotation and history view) to help programmers make sense of collaborators' prompts.
In addition, \sys{} provides four mechanisms for programmers to (1) \textbf{refer} to collaborators' prompts; (2) \textbf{request} intermediate results from collaborators; (3) \textbf{share} information with collaborators; and (4) \textbf{link} variables for synchronization.

\begin{figure*}[h]
    \centering
    \includegraphics[width=\linewidth]{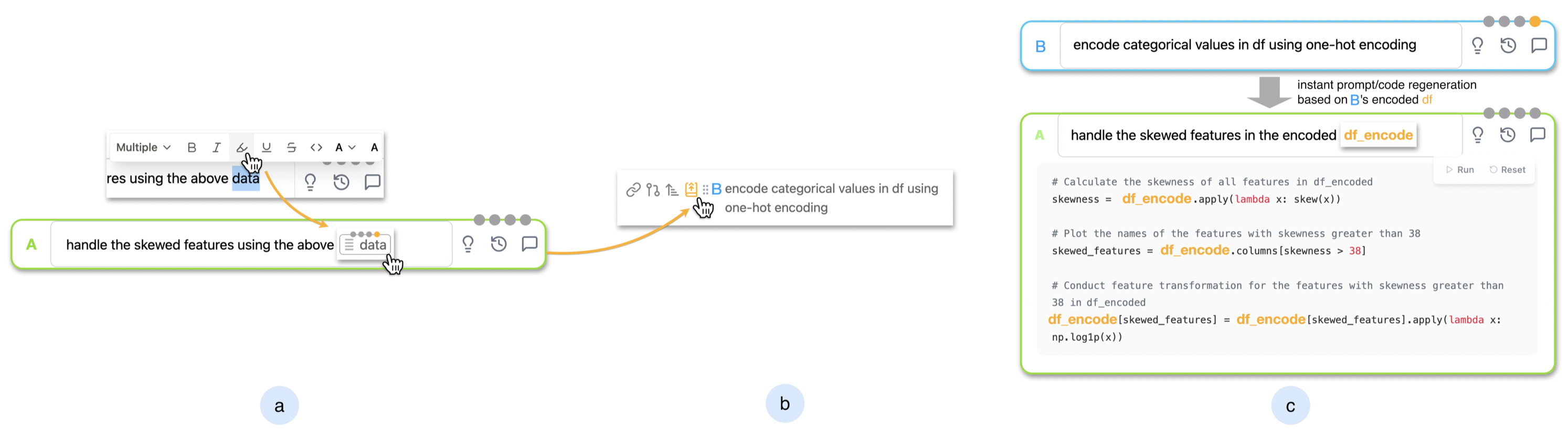}
    \caption{Refer Workflow: (a) select source node from the editor; (b) select target node for reference from the wiki and click the \referlink{} icon; (c) A's block will instantly regenerate prompt and code based on the code and execution result of B's prompt.}
    \Description{There are three parts (a, b, c) of this figure for the workflow of refer: (from left to right): (a) select and highlight prompt utterances that should be modified from the editor as a node; (b) select the target prompt for reference from the wiki as a target node and click the refer icon; (c) A's block will instantly regenerate prompt and code based on the code and execution result of B's prompt.}
    \label{fig:referlink}
\end{figure*}

\subsubsection{Making sense of collaborators' prompts}
In order to leverage collaborators' prompts, programmers first need to make sense of them.
To check the last person who modified the prompt, programmers can check the left of each prompt where the prompt author's icon is displayed.
To assist programmers in comprehending high-level or more complex prompts from collaborators, we designed the \textit{explanations view}
(\textbf{D2}). 
By clicking the \textit{Explain} button (\autoref{fig:UI} d) for the prompt block, programmers can access semantic highlighting to better understand the code structure. 
This feature visually highlights key phrases in both the prompt and the corresponding code segments, linking them to help programmers understand the relationships between them (i.e., which phrase in the prompts led to the generation of a certain line of code).
The view also provides a high-level overview of all the steps within the code, allowing programmers to quickly understand the code's structure, logic, and functionality.
To track the evolution of prompts and their previous versions, programmers can use \textit{history view} to find historical versions of both the prompts and the corresponding code.

In the following sections, we present the design of the four core mechanisms of \sys{}. All four mechanisms contain three components: (1) a \textit{source node} in the prompt blocks that are presented in the editor; (2) a \textit{target node} in the prompt that is listed in the \textit{Prompt Wiki} and (3) additional messages.

\subsubsection{\textbf{Refer} to Collaborators' Prompts for Precise Code Generation}
When programmers need to build upon their collaborators' work, they can use the \referlink{} mechanism (\autoref{fig:referlink}) to specify which part of collaborators' work should be used to provide the context of their prompt (\textbf{D3}).
The programmer needs to first select a \textit{source node} from the editor, which could be a word (e.g., a variable name), a phrase (e.g., perform a function with a variable), or the whole prompt. Upon selection, the editor will toggle the highlighted mark on the selected node.
Next, the programmer chooses a \textit{target node} from the list of prompts displayed in the \textit{Prompt Wiki}. 
This target node could also be a word, a phrase, or the entire prompt. 
Once selected, the \textit{Prompt Wiki} panel will display the corresponding colored icon based on the current actions associated with the node. 
Additionally, a colored dot will be added before the prompt, signifying that this prompt contains nodes with associated actions. Programmers may also opt to include messages to clarify their intentions, thereby improving the accuracy of code generation.

\begin{figure*}[h]
    \centering
    \includegraphics[width=0.99\linewidth]{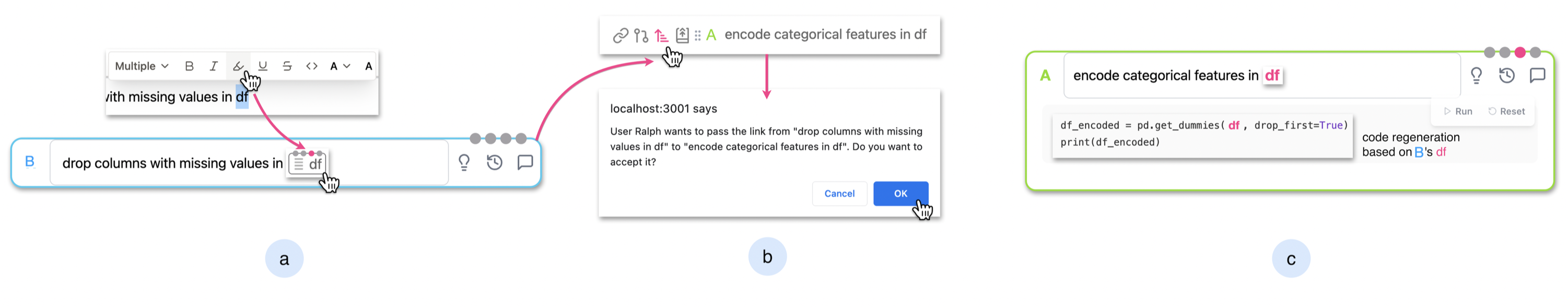}
    \caption{Share Workflow: (a) select a source node to be shared from the editor; (b) select a target node to be updated with the shared content from the wiki and click the \respondlink{} icon to highlight the node, then A will see a pop-up message, indicating that B would like to share some information with A; (c) when A accepts, A's highlighted block will update its code based on B's df.}
    \Description{There are three parts (a, b, c) of this figure for the workflow of share (from left to right): (a) select and highlight the element to be shared as a node from the editor; (b) select the prompt to be updated with the shared content from the wiki as a node and click the share icon to highlight the node, then A will see a pop-up message indicating that the collaborator, B, would like to share some information with A; (c) when A accepts, A's highlighted block will update its code based on B's df.}
    \label{fig:passlink}
\end{figure*}

\subsubsection{\textbf{Request} Collaborators' Assistance for Prompt Engineering}

When programmers require information from their collaborators' incomplete tasks as context for their own prompt, they can follow a simple process with the \textit{Request} feature. First, they create a prompt with a placeholder indicating the expected result from their collaborators. Next, they select a \textit{source node} from the placeholder in the editor and choose a \textit{target node} from the collaborator's work listed in the \textit{Prompt Wiki}. This action notifies the creator of the \textit{target node} that a collaborator is awaiting the task to be resolved. Subsequently, the programmer can continue working on other tasks and return to check on this specific task once the collaborator has finished the requested task (\textbf{D3}).
After creating the request, \sys{} logs and tracks unresolved actions, and when collaborators create prompt blocks that address these actions, it automatically executes the corresponding tasks (\textbf{D2}).

\subsubsection{\textbf{Share} Context to Collaborators}
To share intermediate results or other contexts with collaborators, programmers can employ the \respondlink{} mechanism (\autoref{fig:passlink}). 
After choosing a \textit{source node} in the editor and selecting a \textit{target node} in the \textit{Prompt Wiki}, the collaborator will receive a pop-up notification indicating that shared context is available. Upon accepting the context, the code associated with the prompt that contains the \textit{target node} will be automatically updated with the context and information provided by the source node.
Programmers can also select the \textit{target node} as one of the headings that correspond to a task, especially when there are no existing suitable prompt blocks under that task. In such cases, the recipients can assign this contextual information to the prompt block they create afterward.

\subsubsection{\textbf{Link} Elements for Automatic Synchronization}

In a programming context, a variable, prompt, or code segment within a project may undergo multiple changes and could have various names in blocks authored by different programmers. 
To reduce repetitive updates, programmers can link together variables with the same value but different names. 
By doing so, when the prompt linked to one variable is modified, the associated prompt will automatically update based on the changes. 
This mechanism also extends to variables with procedural dependencies. It not only streamlines future code generation but also helps prevent potential conflicts (\textbf{D4}).

\subsection{System Implementation}
\sys{} is a web-based application built using Next.js and React TypeScript. The core of the block-based editor is constructed using TipTap~\cite{tiptap2023}, which serves as a headless wrapper for ProseMirror~\cite{prosemirror2023}, providing the foundation for a rich text WYSIWYG editor
\rv{
The overall architecture of \sys{} is illustrated in Fig.~\ref{fig:sys-architecture}.

\sys{} enables programmers to begin by creating a prompt block with generated code beneath it. Programmers can also initiate a code block and manually input code, pausing midway to generate a prompt block from the code comments.
\sys{} implements collaborative editing by leveraging the power of Y.js, which is a CRDT-based approach to handling shared document editing. Changes made by users are distributed and merged using WebSocket communication, with the Hocuspocus Server serving as the backend for handling real-time synchronization of documents among users. This approach enables real-time collaboration, syncing between devices, and the ability to work offline while maintaining consistency in the edited documents.
Actions and messages are shared in real-time through the Firebase Real-Time Database~\cite{firebase2023}. The event listener updates the prompt wiki whenever actions are triggered within the four mechanisms.

To facilitate code generation, \sys{} employs the OpenAI GPT-4 API~\cite{openai2023gpt4} in combination with custom prompt templates (all prompt templates are provided in Appendix~\ref{appendix:prompt}). The code generation process leverages the context provided by the prompt wiki, as well as the current prompt and code blocks within block-based text editors.
While \sys{} allows programmers to structure their prompts in any format using any techniques, it is not specifically designed for these techniques (e.g., prompt decomposition or few-shot learning). The primary intention is to enable them to write simple prompts and allow \sys{} to generate the desired code.
In the case of \reciprocatelink{}, \sys{} incorporates automatic self-check mechanisms with a specific prompt template to validate whether changes are necessary when one side is updated.
Regarding the \requestlink{}, all programmers' requests are queued and cross-verified against the expected results specified by requestors. These two mechanisms make use of prompting techniques that draw upon few-shot learning~\cite{brown2020language} and Chain-of-Thought techniques~\cite{wei2023chainofthought} to enhance accuracy in the code generation process.
}

The Python code execution in the web app is made possible through Pyodide~\cite{pyodide2023}. Pyodide represents a port of CPython to WebAssembly, enabling the installation and execution of Python packages directly within the browser using micropip. 
For the execution of Python code in a separate thread, a communication channel is established between the main thread and the Pyodide worker, incorporating a defined communication protocol.
It is worth noting that, despite sharing the same context within the collaborative editor, \sys{} ensures that the OpenAI and Python kernels do not overlap, preserving stability and functionality.

\section{User Study} \label{study}
We conducted a user study to evaluate the effectiveness of \sys{} in assisting programmers' prompt co-engineering during NL programming by answering the following research questions:
\begin{itemize}
    \item \textbf{RQ1}: How does the system support programmers to understand collaborators’ progress and prompts?
    \item \textbf{RQ2}: How does the system support programmers to build on top of collaborators’ work?
    \item \textbf{RQ3}: How does the system minimize the redundant updates of prompts or code?
\end{itemize}

\subsection{Participants and Tasks}
We recruited 12 participants (7 female, 5 male, aged 20-31) from a local university. All participants have more than two months of experience in AI-based code assistants and more than three years of experience in using computational notebooks. Participants were compensated \$50 for the 120-minute study. 
\rv{The data science task was modified from a Kaggle competition~\cite{titanic} that predicts survival from a disaster. To scope the task within the study duration, we asked participants to only perform exploratory data analysis. We also divided the high-level task into several sub-tasks with the assistance of two expert data scientists and provided a basic task division plan for participants' reference.}

\subsection{Procedure}
Participants were first informed of the aim of this study and gave their consent. Then, they were asked to participate in a two-part study including (1) real-time collaboration and (2) following up with the others' work.

\subsubsection{Part 1: Real-time collaboration (90 minutes)}
Participants were asked to perform a data analysis task in pairs using NL prompts collaboratively. Each pair of participants was asked to complete two sessions of collaborative programming: one session to use our \sys{} prototype and another to use the baseline system (VSCode live share with CoPilot plugin). The sequence of the sessions was counter-balanced. 

For each session, the experimenter first introduced features of the system and gave each participant a 10-minute training session on the system, with example tasks to complete.
Then, each pair of participants was asked to work on a data science task collaboratively by prompting for around 30 minutes. After that, participants were asked to rate their experience of \sys{} and the baseline system on a 7-point Likert scale. The experimenters then conducted a semi-structured interview based on the results and observed use patterns to learn participants’ perspectives.

\subsubsection{Part 2: Following up with the Collaboration Process (30 minutes)}
In stage 2, we evaluated how a new collaborator followed up with an ongoing collaborative project using \sys{}. We asked participants to review the work of another pair of participants collected from part 1 of the study using \sys{} (e.g., P1 and P2 reviewed the work of the second pair - P3 and P4). After reviewing the work for 10 minutes, we asked participants to complete additional tasks to modify the existing work, such as changing the way of dealing with outliers. The experimenters then conducted a semi-structured interview regarding the user experience of reviewing and revising others' work using \sys{}.

\subsection{Data Analysis}
All study sessions were recorded and transcribed. Data collection included server-side logs, screen recordings, and interviews. Additionally, we made observational notes during the study. \rv{Our analytical approach involved the articulation of codes and themes, employing a combination of inductive and deductive thematic analysis.
Two authors independently coded the transcripts and identified themes to gain insights into how participants utilized and evaluated \sys{} and the baseline condition. 
The themes generated encompass both parts one and two of the studies, which address the three RQs. We explicitly specified if the results pertain exclusively to part two in the results section.

For all survey data, we opted for non-parametric statistical methods given the ordinal nature of Likert-scale responses and the small sample size. Specifically, we employed the Wilcoxon signed-rank test to compare responses between the two conditions.
Additionally, prompts and code collected from participants' logs underwent open coding based on the reasons for manual modifications.
Two researchers independently open-coded 30\% of the data to establish a codebook, identifying major reasons for prompt and code modifications. These codes were then applied to the remaining data, resulting in a 74\% agreement, which was subsequently refined iteratively until reaching 100\%.
Results from self-defined Likert scale data will be highlighted with question numbers~(Fig.~\ref{fig:survey}).
}
\section{Results}
\begin{figure*}[h]
    \centering
    \includegraphics[width=1\textwidth]{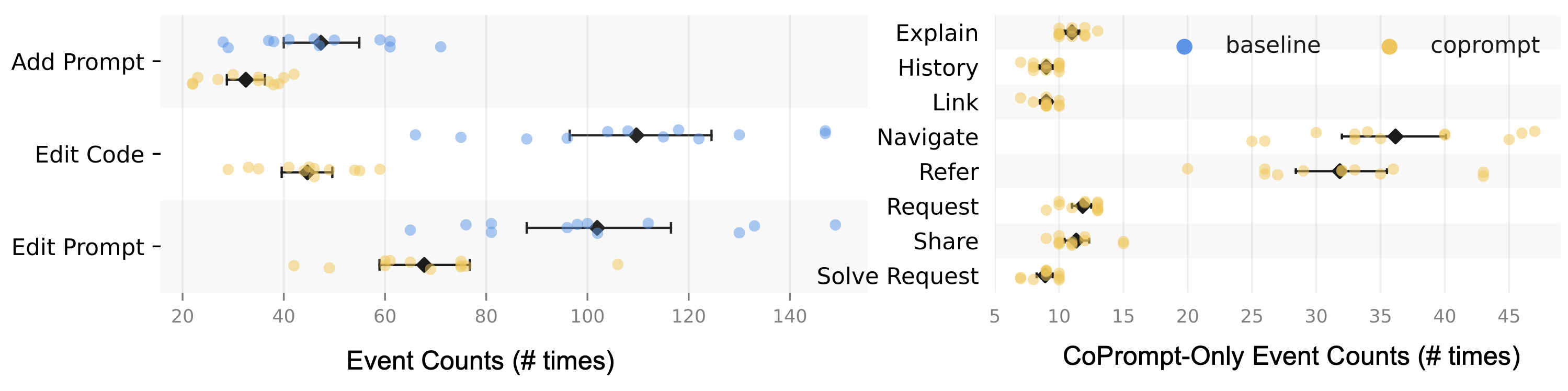}
    \caption{Distribution of event counts per participant. The left image compares the event counts between the baseline and \sys{}, and the right image shows the count of \sys{}-only event types.}
    \Description{Side-by-Side Comparative Visualization of Event Counts: The left panel, marked 'Baseline', exhibits a narrower strip and point plot showcasing the distribution and average counts of event types like 'Add Prompt', 'Edit Code', and 'Edit Prompt'. Alternating grey shading facilitates readability. The right panel, labeled 'Co-prompt', extends wider with a similar plot structure but for a broader array of event types including 'Auto Update', 'Explain', and 'History', among others. Distinctive '\#f6c444' color coding denotes the co-prompt condition, with a legend explicating the hues associated with each condition.}
    \label{fig:event-count}
\end{figure*}

Here we report the findings from analyzing participants’ survey responses, interview transcripts, think-aloud feedback, and system usage logs to understand 1) how \sys{} support prompt co-engineering and 2) how participants perceive the utility of the four mechanisms 
for leveraging collaborators' work and sharing information among collaborators.

\subsection{Overall Collaboration Behaviors and User Perceptions}
\subsubsection{Completion Time}
All participants successfully completed the programming tasks and utilized all four types of mechanisms in the study.
However, participants took significantly more time to complete the tasks in the baseline condition~($M_\sys{}$=$23.21$ < $M\textsubscript{\textit{baseline}}$=$26.82$ minutes, $p$=$.002$, $r$=$.57$). This could be due to heightened synchronous communication requirements in the baseline condition, while in \sys{}, participants leveraged the four mechanisms, facilitating \pqt{a way for quicker communication.}{P3}
\st{Notably, participants engaged in more collaborative workflows when using the \sys{} prototype compared to when using the VSCode editor (baseline). }

\subsubsection{Overall Collaborative Workflow}
In both conditions, all participants initiated their work by crafting prompts at a higher level of abstraction, such as defining the task as \qt{data visualization,} and then iteratively refined these prompts to reach a lower level of detail, like specifying \qt{pair plot df.} 
This process was facilitated by the wiki provided by \sys{}, which allowed participants to easily locate the target prompt block that was available for further iteration. The four mechanisms further supported participants in collaborative efforts without requiring context switching to external communication tools.
P4 explained, \qt{Using \sys{}, I do not need to wait for my collaborator to finish encoding, as I can write prompts for transformation first and then refer to my collaborator's prompt.}
In general, \sys{} facilitated parallel work on programming tasks without being hindered by collaborators' workflows compared to the baseline.

\subsubsection{System Usability \& Cognitive Load}
To measure the usability of \sys{}, we computed the SUS scores based on the UMUX-LITE \cite{Lewis2013UMUX}. The average SUS scores were significantly greater ({\small $p = 0.02$}) for \sys{} ({\small $\text{Mdn} = 90.61$}), compared to baseline ({\small $\text{Mdn} = 68.94$}).
We also used NASA-TLX to measure participants’ perceptions of the cognitive workload of using the systems. Compared to baseline, \sys{} had
lower mental ($\text{Mdn} = 3.5 < 5.5, p = 0.040$), physical ($\text{Mdn} = 1.0 < 3.0, p = 0.0179$), and temporal ($\text{Mdn} = 3.0 < 5.0, p = 0.0082$) demand, required less effort ($\text{Mdn} = 3.0 < 5.5, p = 0.033$), and led to better performance ($\text{Mdn} = 4.5 > 3.0, p = 0.0532$)
and less frustration ($\text{Mdn} = 1.5 < 3, p = 0.0187$).  
The overall perceived workload, obtained by averaging all six raw NASA-TLX scores (with the ``Performance'' measure inverted), was also lower for \sys{} than baseline ($\text{Mdn} = 2.5 < 4, p = 0.0532$).

\subsubsection{Code Edit \& Prompt Edit}
\rv{We compared the overall code and prompt edit counts between the baseline and \sys{} conditions (Figure~\ref{fig:event-count} Left). We observed that while there is a less significant difference in adding prompts ($Mdn_\sys{}$=$35.0$ < $Mdn\textsubscript{\textit{baseline}}$=$46.5$, $p$=$.003$), there are much larger significant differences in code editing ($Mdn_\sys{}$$=45.5$ < $Mdn\textsubscript{\textit{baseline}}$=$111.5$, $p$=$3.42 \times 10^{-8}$) and prompt editing ($Mdn_\sys{}$=$67.0$ < $Mdn\textsubscript{\textit{baseline}}$=$99.0$, $p$=$6.94 \times 10^{-4}$). These substantial differences can be collectively attributed to the four mechanisms and are explained in Sec \ref{subsec:edits}.
}

\subsection{RQ1: How does the system support programmers to understand collaborators’ progress and prompts?}
\sys{} supported programmers' sense-making of collaborators’ work by providing the hierarchical overview, generating explanations associated with the prompts, and displaying historical views. In the following sections, we report the detailed usage and perceived utility of these features. 

\begin{figure*}[h]
    \centering
    \includegraphics[width=0.9\linewidth]{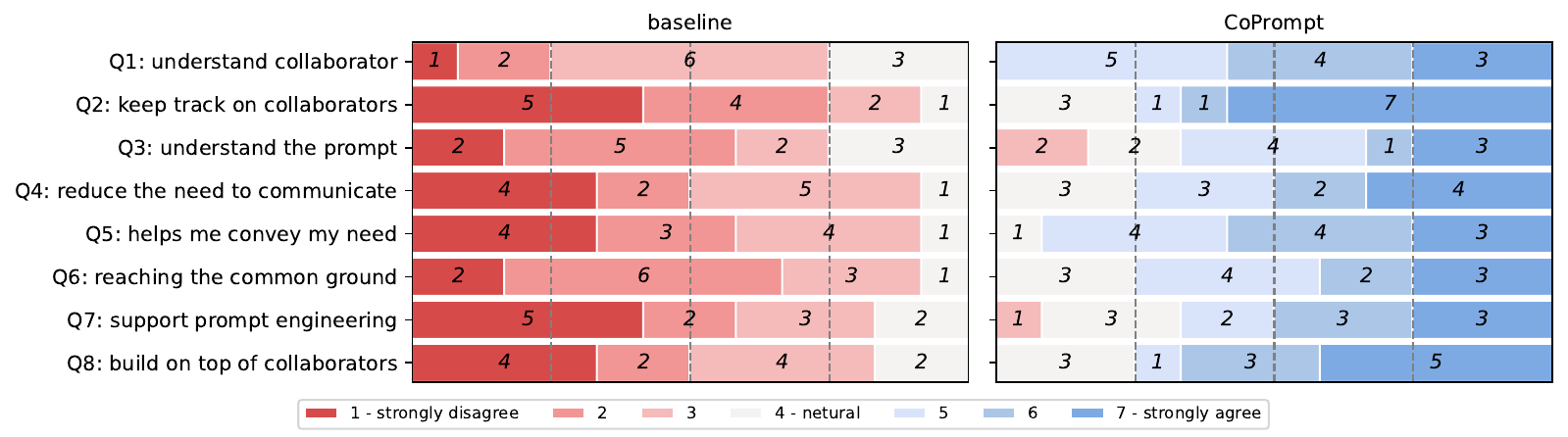}
    \caption{The results from the self-defined Likert scale questionnaire comparing between condition baseline and \sys{}}
    \Description{The image is a bar chart representing the results from a self-defined Likert scale questionnaire comparing two conditions: "baseline" and "CoPrompt." There are eight questions (Q1 through Q8), each corresponding to a different aspect of collaboration or prompt engineering. The responses range from 1 (strongly disagree) to 7 (strongly agree).}
    \label{fig:survey}
\end{figure*}

\subsubsection{The holistic overview with multi-levels of details facilitates users tracking and locating collaborators' processes \textbf{(D1)}.}
All participants found that the prompt wiki reduces the need for programmers to keep track of collaborators' progress (Q2: $\text{Mdn} = 6.5 > 2, p = 0.0034$, \rv{\autoref{fig:survey}}) and helps them understand what the collaborators are doing (Q1: $\text{Mdn} = 5.5 > 3, p = 0.0034$). 
The foldable table-of-content-like view allowed participants to view all the sections of the notebook at a higher level, which facilitated participants in locating their target regions and \pqt{understand the overall structure easily.}{P4}
With the navigation function, participants could navigate to their target regions by clicking on the items in the prompt wiki, which is straightforward as \pqt{there is no need to scroll back and forth.}{P1} 
\rv{The event count from the log data~(Figure~\ref{fig:event-count} Right) shows that participants frequently utilize this navigation feature ($M_\sys{}$=$36.17$, $SD$=$7.47$).}
P2 expressed their preference for the annotation in the wiki as it indicated the owner of the prompt: \qt{With the icon before the prompt, the ownership of the prompt is clear.} 
We also noticed that most participants (N=9) collapsed their own sections while leaving the collaborator's sections expanded to \pqt{track on collaborators' work}{P4} and \pqt{easily identify changes not made by me [participant].}{P11}
\rv{In part two of the study, participants utilize the prompt wiki to gain a quick understanding of the asynchronous work accomplished by their collaborators. 
They subsequently used the navigation feature to look into lower-level code details when they encountered \pqt{tasks that were unclear.}{P5}}

\subsubsection{Generating and associating explanations assisted participants' comprehension of prompts \textbf{(D2)}.} 
All participants found that the system helps them understand the prompt written by collaborators (Q3: $\text{Mdn} = 5 > 2.5, p = 0.0304$). They mentioned that the high-level, step-by-step explanations of the generated code and the semantic connections between prompts and code segments
are \pqt{especially useful when the prompt or the code is way too long.}{P1} 
Participants typically required these explanations before utilizing the \textbf{refer} mechanism, as sometimes they struggled to fully grasp their collaborators' intentions through the prompts alone. The prompt explanations played a vital role in scaffolding participants' comprehension by \pqt{providing more details about the code.}{P1} Consequently, participants could reuse or refer to their collaborators' work without the need for direct communication, as highlighted by one participant, \pqt{I do not need to keep bothering my partner to ask what the code is about.}{P9}
Furthermore, some participants perceived these explanations as a means to reduce conflicts and errors during collaboration. As one participant noted, \pqt{I became less likely to misunderstand my collaborators' intention and modify their code, which always causes conflicts.}{P1}

\subsubsection{Providing a history view allows users to have a clear view of the changes \textbf{(D1)}.} 
The history view presents a comprehensive record of all historical versions of a specific prompt, offering users insight into its evolving process, as noted by P5, \qt{It is good to view the changes.} This view not only captures the modifications made but also the \pqt{interactions that happened}{P1} and the \pqt{generated result for each iteration}{P2}, thus facilitating a holistic understanding of the prompt's evolution.
Participants frequently utilized this feature when changes were enacted after any mechanism had been activated (e.g., the requested message had been resolved). In such cases, the history view allowed them to identify \pqt{who altered my code.}{P2} 
Most participants (N=8) also leveraged the diff view of the code and prompt to resolve the conflict and restore the version. Overall, participants reported that the system supports programmers in reaching common ground with collaborators ($\text{Mdn} = 5 > 2, p = 0.0034$).
Interestingly, participants also employed the history view to check if their collaborators had \pqt{started work on my requests.}{P12} 
Moreover, the historical view also helped participants refine their prompts and serve as memory anchors. For instance, P7 mentioned, \qt{it helped me recall what or why I made certain changes.}

\subsection{RQ2: How does the system support programmers to leverage collaborators’ work?}
All participants agreed that the four types of mechanisms are 
\textbf{(1) easy to use:} \referlink{} (\rv{$Mdn$=$6.5$}, $SD$=$1.76$), \requestlink{} ($Mdn$=$5.5$, $SD$=$1.35$), \respondlink{} ($Mdn$=$6$, $SD$=$1.45$) and \reciprocatelink{} ($Mdn$=$6$, $SD$=$1.27$);
\textbf{(2) intuitive to learn:} \referlink{} ($Mdn$=$6$, $SD$=$1.15$), \requestlink{} ($Mdn$=$5$, $SD$=$1.68$), \respondlink{} ($Mdn$=$5.5$, $SD$=$1.31$) and \reciprocatelink{} ($Mdn$=$6$, $SD$=$1.30$);
\textbf{(3) could fulfill their requirements:} \referlink{} ($Mdn$=$6$, $SD$=$1.68$), \requestlink{} ($Mdn$=$5.5$, $SD$=$1.53$), \respondlink{} ($Mdn$=$6$, $SD$=$1.61$) and \reciprocatelink{} ($Mdn$=$6$, $SD$=$1.36$);
\rv{and \textbf{(4) easy to control:} \referlink{} ($Mdn$=$6$, $SD$=$1.37$), \requestlink{} ($Mdn$=$5$, $SD$=$1.03$), \respondlink{} ($Mdn$=$5$, $SD$=$1.61$) and \reciprocatelink{} ($Mdn$=$5.5$, $SD$=$1.51$).}
Overall, \sys{} reduces participants' need to communicate with collaborators (Q4: $\text{Mdn} = 5 > 3, p = 0.0122$) and facilitates participants to modify their prompt (Q7: $\text{Mdn} = 5.5 > 2.5, p = 0.0065$).

\subsubsection{Request and auto-updates reduce mental load \textbf{(D3)}}
\sys{} helps participants to convey their needs clearly to the collaborator (Q5: $\text{Mdn} = 5.5 > 2.5, p = 0.0122$) and reduce the need to \pqt{remember what tasks have not yet been done.}{P9}
In collaborative programming, since the task distribution may involve procedural dependencies, there are often cases where participants need to wait for their collaborators to complete certain tasks. 
With the request-detect-update mechanism, participants just need to send request with brief descriptions.
\rv{Participants reported that although it \pqt{took more time to get familiar with [\requestlink{}]}{P10}, it ultimately saved them considerable time.}
They could request tasks at higher-level headings if a suitable prompt had not yet been created by collaborators.

\begin{figure*}[h]
    \centering
    \includegraphics[width=1\linewidth]{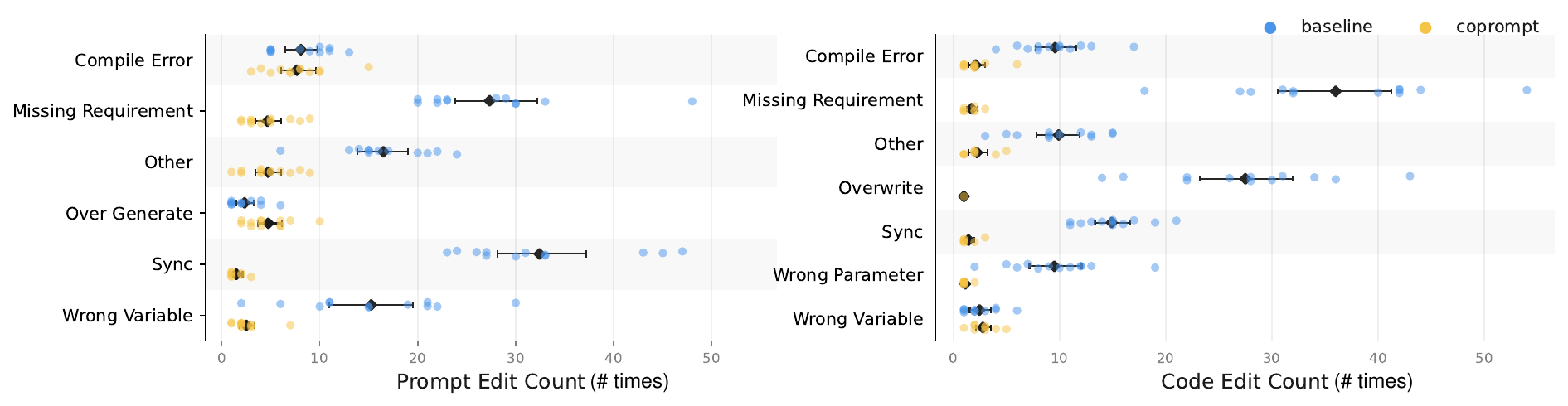}
    \caption{The left image displays edit prompt counts categorized by reasons of change for both the baseline and \sys{} conditions; The right image illustrates code edit counts categorized by reasons of change for the same two conditions.}
    \label{fig:edit-reason}
    \Description{The visualization shows the edit prompt and code edit counts categorized by reasons, comparing the "baseline" and "coprompt" conditions. In the edit prompt section (left), there is a clear distinction between the reasons "Compile Error," "Missing Requirement," "Other," "Over Generate," "Sync," and "Wrong Variable" for both conditions. In the code edit section (right), the reasons "Compile Error," "Missing Requirement," "Other," "Overwrite," "Sync," "Wrong Parameter," and "Wrong Variable" are compared between the two conditions.}
\end{figure*}

Compared to communicating with collaborators using messages or audio in baseline, the request takes less cognitive effort since \pqt{it provides contextual information through the node}{P3} and does not request programmers to \pqt{write too much text to describe the changes.}{P2}
Some participants (N=4) highlighted that this feature reduces cognitive load of ensuring a polite tone, as they no longer need to carefully craft messages to their collaborators, \pqt{I do not need to care about the manner.}{P3}
In addition, the auto-update of the prompts sending requests after target input is detected from collaborators offloads the tedious review and update process. P9 and P10 expressed their satisfaction with these features: \pqt{Detection and auto-update saved me a lot of effort.}{P10} 
P9 added, \qt{I only need to know what the task is about, without thinking about where and when to address it.} 
However, the automatic update feature makes P12 feel a slight loss of control, and sometimes they would like to decide whether to handle updates manually or automatically.

\subsubsection{Sharing knowledge both proactively and reactively \textbf{(D4)}}
Participants utilized the \respondlink{} mechanism for both proactive and reactive sharing of output or prompt segments. Proactively, some participants shared their results with collaborators, \pqt{I know the encoded result would be used for following step.}{P1} They also shared insights from the result through \respondlink{}, \pqt{I would share some text that may not necessary is the prompt to share insights I got.}{P7} Reactively, participants responded to their collaborators' requests for specific data processing results through direct communication by sharing requested results. Participants also made use of the hierarchical structure to share elements by placing them under relevant headings. This occurred when collaborators had not yet delved into those specific details.

For the receiver of the shared content, most participants indicated that the content was useful for clarifying their own prompts and the pop-up message was clear enough for quick comprehension, \pqt{very convenient as I do not need to find and refer to my collaborators' work.}{P5} Participants also revealed their trial-and-error strategies for dealing with the pop-up message that is too brief to understand or too long to read, \pqt{Just like using ChatGPT, I just accept the answer and view its execution result to evaluate its effect.}{P3} This strategy was adopted by many participants (N=9), as the trial-and-error cost is low with the view of the history version and the \pqt{ability to trace back to previous versions.}{P4}

\subsubsection{Referencing reduces the effort of careful reading and copy-pasting \textbf{(D4)}}
\sys{} helps participants build on top of collaborators' work easily (Q8: $\text{Mdn} = 6 > 3, p = 0.0049$). 
All participants used the \referlink{} feature significantly more than \rv{other links when using \sys{} to perform prompt co-engineering (Figure~\ref{fig:event-count} Right).
} 
With the \referlink{} mechanism, participants no longer need to read the whole prompt and select utterances for copy-pasting to modify their own prompts. Instead, they handed off the comprehension work to \sys{} by guiding it with the \referlink{} mechanism. 
The multi-level hierarchy display of prompts supported programmers in pinpointing the prompts they wished to refer to and enabled them to select the most appropriate level for reference.
All participants agreed that the interaction process of \referlink{} \pqt{reduced the cognitive switching between communication and code.}{P5}

\rv{Analyzing the open-coded results regarding the reasons for manually modifying prompts and code~(Fig.~\ref{fig:edit-reason}), we observed that participants using \sys{} made significantly fewer modifications due to \qt{Missing Requirements} in both code and prompt edits (Code Edit: $Mdn_\sys{}$=$2$ < $Mdn\textsubscript{\textit{baseline}}$=$36$, $p$<$5.10 \times 10^{-8}$; Prompt Edit: $Mdn_\sys{}$=$4.5$ < $Mdn\textsubscript{\textit{baseline}}$=$25.5$, $p$=$2.57 \times 10^{-9}$). 
Similarly, we found that participants in the baseline condition needed to make significantly more modifications to the prompt and code due to \qt{Wrong Variable}, which was caused by outdated or incorrect variables generated by the AI model (Code Edit: $Mdn_\sys{}$=$2$ < $Mdn\textsubscript{\textit{baseline}}$=$9.5$, $p$=$1.93 \times 10^{-5}$; Prompt Edit: $Mdn_\sys{}$=$2$ < $Mdn\textsubscript{\textit{baseline}}$=$15$, $p$=$1.39 \times 10^{-5}$).
While these results can be attributed to all four mechanisms that collectively reduce the overall need for manual modifications, the substantial reduction in the need to modify due to missing requirements and wrong variables is more likely a result of the mechanism \referlink{}.
}

\subsection{RQ3: How does the system reduce the repetitive updates of prompts or code?} \label{subsec:edits}

The \reciprocatelink{} mechanism effectively reduced the frequency of repetitive updates by offering automatic synchronization \textbf{(D2)}.
\rv{Participants made significantly fewer modifications to code and prompts due to \qt{Sync} when using \sys{} ($Mdn_\sys{}$=$1$ < $Mdn\textsubscript{\textit{baseline}}$=$28$, $p$<$.001$), indicating a reduced need to update in response to changes made by others~(Fig.~\ref{fig:edit-reason}).}
\sys{} decreased the need for participants to repeatedly and iteratively modify prompts and significantly facilitated their ability to establish a shared understanding with collaborators (Q6: $\text{Mdn} = 5 > 2, p = 0.0068$).

The \reciprocatelink{} mechanism was deemed the most intuitive by all participants, and they unanimously agreed that it significantly reduced the workload associated with repetitive updates due to procedural dependencies. P4 expressed that it \qt{saved a lot of time on going back and forth between users} and \pqt{helped offload some mental model}{P5} without the need to keep track of collaborators' changes on a certain task. Additionally, three participants employed links between headings to convey the synchronization of an entire subsection.

All participants mentioned that the \referlink{} and \respondlink{} mechanism simplifies the interaction of updating prompts.
When there is a need to update existing prompts, many participants (N=9) leveraged \referlink{} so that they just needed to indicate the part of the prompt that requires modification and the target prompt reference.
The \respondlink{} mechanism assisted programmers' prompt engineering by allowing collaborators to pass knowledge to others and it only requires receivers to accept the shared prompt or code snippets and coarsely navigate to the target prompt, instead of \pqt{carefully locating and manual copy-pasting.}{P7}
\rv{However, it also increases the likelihood of the models generating excessive content (as shown in Figure~\ref{fig:edit-reason} left), requiring participants to manually refine the prompts ($Mdn_\sys{}$=$3$ > $M\textsubscript{\textit{baseline}}$=$2$, $p$=$.63$).}
\section{Discussion}
We discuss various topics that emerged during our study, offering insights and design implications. We further outline the primary limitations of our study and highlight areas for future research.

\subsection{Using NL-predominant Expression in Collaborative Programming}
\rv{
Compared to code, NL-predominant expressions are easier to understand~\cite{Miller1981Natural} and therefore helpful for maintaining shared understanding among collaborators~\cite{Pang2022How, Padioleau2009Listening}. 
However, programmers still have difficulty making sense of the NL expressions used by their collaborators when they are too vague or at a high level. CoPrompt provided the hierarchical wiki, explanation view, and history view to assist programmers' comprehension with different levels of details. 
CoPrompt also allows programmers to share knowledge without losing context through four mechanisms, reducing their need for repetitive updates, copy-pasting, and synchronization in collaboration.
Our findings reveal the possible benefits of prompt sharing and referring, including reducing task completion time, the need for communication and cognitive load. Participants found it useful to share intermediate results with context like the method details. However, it also increased the likelihood of the models generating excessive content, requiring participants to manually refine it.
}

\subsection{Sense of Control in Code Generation Process}
\rv{
In \sys{}, the four mechanisms involve a high level of automation \cite{mackeprang2019sweetSpot} as LLMs are in charge of comprehending. This design significantly reduced the programmers' cognitive load, while it raised concerns about the generated content: How to make the generation more satisfying and how to efficiently deal with unsatisfying results.
Prior work in the domain of human-AI collaboration emphasized the importance of preserving user control when collaborating with LLMs~\cite{Zhang2023VISAR} at different granularities~\cite{Wu2022AI}. 
To maintain user control in code generation process, \sys{} allows programmers to manually tweak the code whenever the result is unsatisfying, and provides convenient history views for each prompt for checking and rolling back. 
Our design serves as a first step towards interaction design in prompt co-engineering, using cases of simple prompting templates to demonstrate the effectiveness of the workflow and core mechanisms.
Future work could extend the design by considering more complex prompting strategies like few-shot prompts. The level of generation automation and controllability provided to the programmers could also be further investigated.
}

\subsection{Synchronous and Asynchronous Collaboration}
\rv{Overall, \sys{} can be used in both synchronous and asynchronous collaboration settings. While the overall design of \sys{} is catered towards synchronous programming, some features could assist asynchronous tasks. For instance, the message panel stores and displays each programmer's usage of the mechanisms, which facilitates programmers who are initially offline to catch up with their collaborators' work.
Although the formative study was conducted in real-time settings, 
participants overall followed the scatter-gather interaction~\cite{Wang2019Human}, where they did not synchronously work on the task.
Many interactions and communications were not carried out in real-time,
where participants worked on their distributed tasks independently and then gathered the results without close and back-and-forth communication. All participants spent more than 15 minutes working asynchronously.
We also evaluated the usage of \sys{} in both synchronous and asynchronous settings using the 2-part user study: part 1 as real-time collaboration and part 2 as asynchronous collaboration.
This demonstrates that \sys{} is able to support both sync and async collaborative programming.
However, there are some specific challenges for asynchronous collaboration that the current \sys{} could not fully tackle, such as preserving essential elements when automatically updating offline collaborators' work. Future work could investigate ways of balancing user control and level of automation. 
}

{\subsection{Supporting Mixed Collaborative Programming Styles}}

\rv{
Although \sys{} aims to facilitate prompt co-engineering, it does not depend on the ``NL-first'' style of collaboration. It also supports code-level collaboration, which allows programmers to manually write and modify code in the generation block.
If programmers would like to have multiple prompts and code in one block, they can write multiple prompts inside the block, each starting a new line. The code will then be generated exactly below the prompts.

Participants edit both prompts and code when using \sys{} (Figure~\ref{fig:event-count}), but they tend to add prompt blocks more frequently than code blocks.
We observed that three participants opted to \emph{add} new code blocks directly and used the four mechanisms on the code itself. They explained that this approach provided them with greater \pqt{controllability}{P9} over their program.
Additionally, we observed several differences between NL-first and code-level collaboration when participants interacted with \sys{}. 1) NL prompts served as high-level summaries, while code sharing often involved larger code chunks that required more navigation; 2) NL prompts were generally declarative and conveyed meaning directly, though exceptions existed for participants who added code blocks when code was in one line and \pqt{declarative}{P5} enough; 3) Participants' familiarity with programming tasks also influenced their preference. Those less familiar with data science preferred prompts that mostly explain the intentions behind them. Future research could investigate the differences in using \sys{} with NL and code. We envision the possibility of a mixed-methods approach that combines elements of code and NL for more efficient collaboration.

Further, all four mechanisms in \sys{} can be generalized to code-level manipulation by selecting code snippets as nodes.
For instance, when utilizing the \reciprocatelink{} to link two variables in code, the code will automatically update based on the context whenever the user-specified condition is met. This feature is particularly valuable in programming languages plagued by issues related to procedural dependencies~\cite{Liu2021Boba}.
However, the effectiveness of the \respondlink{} mechanism might be limited at the code level. When a programmer wants to share a piece of code with collaborators, merely sharing the code without prompt might result in misinterpretation for both the LLM and the collaborator.
Future research could distinguish between code and prompt sharing in collaborative NL programming.

}

\subsection{Limitations}
\rv{Our study of the \sys{} presents important findings in the domain of collaborative NL programming. However, certain limitations should be recognized.
First, CoPrompt was exclusively tested with programmers experienced in using LLM-driven code assistants. While NL programming is beneficial to a wide range of expertise levels,
our pilot study indicated that programmers with limited experience often found it difficult to refine prompts effectively and were less active in the collaborative process. Future studies should be conducted including programmers unfamiliar with AI-driven code assistants
to understand how varying expertise levels influence the 
collaborative workflows.

Secondly, we chose data science work as our case study due to its involvement of diverse participants~\cite{Crisan2021Passing} and the existence of procedural dependencies among various artifacts~\cite{Liu2021Boba, Sarma2023Multiverse}. The exploratory and explanatory aspects of data science necessitate close collaboration, frequent information exchange~\cite{Chattopadhyay2020what, Liu2020Paths, pang2022data}, and discussions~\cite{Wang2020Callisto}. These characteristics made data science an ideal initial case to explore the concept of prompt co-engineering.
Beyond data science, challenges such as repetitive updating and synchronization persist~\cite{Ying2020Understanding} in general collaborative programming~\cite{Williams2000Strengthening}. 
Since the workflow and mechanisms of \sys{} are designed for prompt co-engineering rather than specific data science tasks, they remain applicable to reduce the need for repetitive updating and the effort for synchronization.
Nonetheless, in certain cases, 
there might be some challenges that the current \sys{} could not fully tackle, such as dealing with compilation error~\cite{Goldman2011Real} and organizing spaghetti code~\cite{Myers1991Separating}. For instance, \sys{} could automatically update some linked artifacts in the spaghetti code, but could not ensure that there are no conflicts due to the entangled code structure.
To better cope with the specific challenges in the wider domain of collaborative programming, future work should incorporate more customized designs.

While \sys{} leverages a rich text editor with block-formatting features, it also works for less structured files or projects spanning multiple files, which are common in general programming tasks besides data science.
The coding blocks used in \sys{} are the objectification of code snippets and prompt phrases, leading the prompt co-engineering to a node-based workflow. By selecting any part of the project as a node, all mechanisms could be applied to the selected node. 
Meanwhile, the four mechanisms enabled participants to distribute the tasks into smaller sub-tasks. 
Prior work mainly investigated the collaborative styles of single authoring, divide \& conquer and competitive authoring \cite{Wang2019How, Posner1992How}. 
Our proposed mechanisms could lower the cost of merging task results and leveraging others' output. This could enable the task distribution to be more nuanced so that every collaborator could program simultaneously, resulting in a more parallel collaboration and thus higher efficiency. 
This may also prevent the formation of spaghetti code.
}

\section{Conclusion}
In this work, we investigated the potential workflow of using NL prompts to conduct collaborative programming, especially prompt co-engineering. Our formative study revealed the workflow of prompt co-engineering and identified four challenges related to comprehension, synchronization, feedback, and reference of collaborators' prompts. To tackle these challenges, we introduced \sys{}, a prototype to support prompt co-engineering through four novel mechanisms: share, refer, request, and link. Our user study indicated that \sys{} effectively supported programmers' workflow of prompt co-engineering from comprehending collaborators' work to leveraging and sharing work.

\begin{acks}
This research was partially supported by the 2021 CCF-Tencent Rhino-Bird Research Fund, the Research Matching Grant Scheme (RMGS, Project No. 9229095), 2023 Guangzhou Science and Technology Program City-University Joint Funding Project (Project No. 2023A03J0001), Guangdong Provincial Key Lab of Integrated Communication, Sensing and Computation for Ubiquitous Internet of Things (No.2023B1212010007), and Natural Sciences and Engineering Research Council of Canada (NSERC) Discovery Grant.
We thank our reviewers for their constructive feedback and participants for their participation.

\end{acks}

\bibliographystyle{ACM-Reference-Format}
\bibliography{main}


\begin{thebibliography}{96}


\ifx \showCODEN    \undefined \def \showCODEN     #1{\unskip}     \fi
\ifx \showDOI      \undefined \def \showDOI       #1{#1}\fi
\ifx \showISBNx    \undefined \def \showISBNx     #1{\unskip}     \fi
\ifx \showISBNxiii \undefined \def \showISBNxiii  #1{\unskip}     \fi
\ifx \showISSN     \undefined \def \showISSN      #1{\unskip}     \fi
\ifx \showLCCN     \undefined \def \showLCCN      #1{\unskip}     \fi
\ifx \shownote     \undefined \def \shownote      #1{#1}          \fi
\ifx \showarticletitle \undefined \def \showarticletitle #1{#1}   \fi
\ifx \showURL      \undefined \def \showURL       {\relax}        \fi
\providecommand\bibfield[2]{#2}
\providecommand\bibinfo[2]{#2}
\providecommand\natexlab[1]{#1}
\providecommand\showeprint[2][]{arXiv:#2}

\bibitem[Git(2023)]%
        {Github2022Copilot}
 \bibinfo{year}{2023}\natexlab{}.
\newblock \bibinfo{title}{Copilot: Your AI pair programmer}.
\newblock
\newblock
\urldef\tempurl%
\url{https://github.com/features/copilot}
\showURL{%
\tempurl}


\bibitem[Anna~Montoya(2016)]%
        {houseprices}
\bibfield{author}{\bibinfo{person}{DataCanary Anna~Montoya}.} \bibinfo{year}{2016}\natexlab{}.
\newblock \bibinfo{title}{House Prices - Advanced Regression Techniques}.
\newblock
\newblock
\urldef\tempurl%
\url{https://kaggle.com/competitions/house-prices-advanced-regression-techniques}
\showURL{%
\tempurl}


\bibitem[Beurer-Kellner et~al\mbox{.}(2023)]%
        {Beurer2023Prompting}
\bibfield{author}{\bibinfo{person}{Luca Beurer-Kellner}, \bibinfo{person}{Marc Fischer}, {and} \bibinfo{person}{Martin Vechev}.} \bibinfo{year}{2023}\natexlab{}.
\newblock \showarticletitle{Prompting Is Programming: A Query Language for Large Language Models}.
\newblock \bibinfo{journal}{\emph{Proc. ACM Program. Lang.}} \bibinfo{volume}{7}, \bibinfo{number}{PLDI}, Article \bibinfo{articleno}{186} (\bibinfo{date}{jun} \bibinfo{year}{2023}), \bibinfo{numpages}{24}~pages.
\newblock
\urldef\tempurl%
\url{https://doi.org/10.1145/3591300}
\showDOI{\tempurl}


\bibitem[B{\o}dker(2015)]%
        {bodker2015third}
\bibfield{author}{\bibinfo{person}{Susanne B{\o}dker}.} \bibinfo{year}{2015}\natexlab{}.
\newblock \showarticletitle{Third-wave HCI, 10 years later---participation and sharing}.
\newblock \bibinfo{journal}{\emph{interactions}} \bibinfo{volume}{22}, \bibinfo{number}{5} (\bibinfo{year}{2015}), \bibinfo{pages}{24--31}.
\newblock


\bibitem[Braun and Clarke(2012)]%
        {braun2012thematic}
\bibfield{author}{\bibinfo{person}{Virginia Braun} {and} \bibinfo{person}{Victoria Clarke}.} \bibinfo{year}{2012}\natexlab{}.
\newblock \bibinfo{booktitle}{\emph{Thematic analysis.}}
\newblock \bibinfo{publisher}{American Psychological Association}.
\newblock


\bibitem[Brown et~al\mbox{.}(2020)]%
        {brown2020language}
\bibfield{author}{\bibinfo{person}{Tom Brown}, \bibinfo{person}{Benjamin Mann}, \bibinfo{person}{Nick Ryder}, \bibinfo{person}{Melanie Subbiah}, \bibinfo{person}{Jared~D Kaplan}, \bibinfo{person}{Prafulla Dhariwal}, \bibinfo{person}{Arvind Neelakantan}, \bibinfo{person}{Pranav Shyam}, \bibinfo{person}{Girish Sastry}, \bibinfo{person}{Amanda Askell}, {et~al\mbox{.}}} \bibinfo{year}{2020}\natexlab{}.
\newblock \showarticletitle{Language models are few-shot learners}.
\newblock \bibinfo{journal}{\emph{Advances in neural information processing systems}}  \bibinfo{volume}{33} (\bibinfo{year}{2020}), \bibinfo{pages}{1877--1901}.
\newblock


\bibitem[Chattopadhyay et~al\mbox{.}(2020)]%
        {Chattopadhyay2020what}
\bibfield{author}{\bibinfo{person}{Souti Chattopadhyay}, \bibinfo{person}{Ishita Prasad}, \bibinfo{person}{Austin~Z. Henley}, \bibinfo{person}{Anita Sarma}, {and} \bibinfo{person}{Titus Barik}.} \bibinfo{year}{2020}\natexlab{}.
\newblock \showarticletitle{What's Wrong with Computational Notebooks? Pain Points, Needs, and Design Opportunities}. In \bibinfo{booktitle}{\emph{Proceedings of the 2020 CHI Conference on Human Factors in Computing Systems}} (Honolulu, HI, USA) \emph{(\bibinfo{series}{CHI '20})}. \bibinfo{publisher}{Association for Computing Machinery}, \bibinfo{address}{New York, NY, USA}, \bibinfo{pages}{1–12}.
\newblock
\showISBNx{9781450367080}
\urldef\tempurl%
\url{https://doi.org/10.1145/3313831.3376729}
\showDOI{\tempurl}


\bibitem[Chen et~al\mbox{.}(2021)]%
        {chen2021evaluating}
\bibfield{author}{\bibinfo{person}{Mark Chen}, \bibinfo{person}{Jerry Tworek}, \bibinfo{person}{Heewoo Jun}, \bibinfo{person}{Qiming Yuan}, \bibinfo{person}{Henrique~Ponde de Oliveira~Pinto}, \bibinfo{person}{Jared Kaplan}, \bibinfo{person}{Harri Edwards}, \bibinfo{person}{Yuri Burda}, \bibinfo{person}{Nicholas Joseph}, \bibinfo{person}{Greg Brockman}, \bibinfo{person}{Alex Ray}, \bibinfo{person}{Raul Puri}, \bibinfo{person}{Gretchen Krueger}, \bibinfo{person}{Michael Petrov}, \bibinfo{person}{Heidy Khlaaf}, \bibinfo{person}{Girish Sastry}, \bibinfo{person}{Pamela Mishkin}, \bibinfo{person}{Brooke Chan}, \bibinfo{person}{Scott Gray}, \bibinfo{person}{Nick Ryder}, \bibinfo{person}{Mikhail Pavlov}, \bibinfo{person}{Alethea Power}, \bibinfo{person}{Lukasz Kaiser}, \bibinfo{person}{Mohammad Bavarian}, \bibinfo{person}{Clemens Winter}, \bibinfo{person}{Philippe Tillet}, \bibinfo{person}{Felipe~Petroski Such}, \bibinfo{person}{Dave Cummings}, \bibinfo{person}{Matthias Plappert}, \bibinfo{person}{Fotios Chantzis},
  \bibinfo{person}{Elizabeth Barnes}, \bibinfo{person}{Ariel Herbert-Voss}, \bibinfo{person}{William~Hebgen Guss}, \bibinfo{person}{Alex Nichol}, \bibinfo{person}{Alex Paino}, \bibinfo{person}{Nikolas Tezak}, \bibinfo{person}{Jie Tang}, \bibinfo{person}{Igor Babuschkin}, \bibinfo{person}{Suchir Balaji}, \bibinfo{person}{Shantanu Jain}, \bibinfo{person}{William Saunders}, \bibinfo{person}{Christopher Hesse}, \bibinfo{person}{Andrew~N. Carr}, \bibinfo{person}{Jan Leike}, \bibinfo{person}{Josh Achiam}, \bibinfo{person}{Vedant Misra}, \bibinfo{person}{Evan Morikawa}, \bibinfo{person}{Alec Radford}, \bibinfo{person}{Matthew Knight}, \bibinfo{person}{Miles Brundage}, \bibinfo{person}{Mira Murati}, \bibinfo{person}{Katie Mayer}, \bibinfo{person}{Peter Welinder}, \bibinfo{person}{Bob McGrew}, \bibinfo{person}{Dario Amodei}, \bibinfo{person}{Sam McCandlish}, \bibinfo{person}{Ilya Sutskever}, {and} \bibinfo{person}{Wojciech Zaremba}.} \bibinfo{year}{2021}\natexlab{}.
\newblock \bibinfo{title}{Evaluating Large Language Models Trained on Code}.
\newblock
\newblock
\showeprint[arxiv]{2107.03374}~[cs.LG]


\bibitem[Chen et~al\mbox{.}(2017)]%
        {Chen2017Codeon}
\bibfield{author}{\bibinfo{person}{Yan Chen}, \bibinfo{person}{Sang~Won Lee}, \bibinfo{person}{Yin Xie}, \bibinfo{person}{YiWei Yang}, \bibinfo{person}{Walter~S. Lasecki}, {and} \bibinfo{person}{Steve Oney}.} \bibinfo{year}{2017}\natexlab{}.
\newblock \showarticletitle{Codeon: On-Demand Software Development Assistance}. In \bibinfo{booktitle}{\emph{Proceedings of the 2017 CHI Conference on Human Factors in Computing Systems}} (Denver, Colorado, USA) \emph{(\bibinfo{series}{CHI '17})}. \bibinfo{publisher}{Association for Computing Machinery}, \bibinfo{address}{New York, NY, USA}, \bibinfo{pages}{6220–6231}.
\newblock
\showISBNx{9781450346559}
\urldef\tempurl%
\url{https://doi.org/10.1145/3025453.3025972}
\showDOI{\tempurl}


\bibitem[Chowdhary and Chowdhary(2020)]%
        {chowdhary2020natural}
\bibfield{author}{\bibinfo{person}{KR1442 Chowdhary} {and} \bibinfo{person}{KR Chowdhary}.} \bibinfo{year}{2020}\natexlab{}.
\newblock \showarticletitle{Natural language processing}.
\newblock \bibinfo{journal}{\emph{Fundamentals of artificial intelligence}} (\bibinfo{year}{2020}), \bibinfo{pages}{603--649}.
\newblock


\bibitem[Crisan et~al\mbox{.}(2021)]%
        {Crisan2021Passing}
\bibfield{author}{\bibinfo{person}{Anamaria Crisan}, \bibinfo{person}{Brittany Fiore-Gartland}, {and} \bibinfo{person}{Melanie Tory}.} \bibinfo{year}{2021}\natexlab{}.
\newblock \showarticletitle{Passing the Data Baton : A Retrospective Analysis on Data Science Work and Workers}.
\newblock \bibinfo{journal}{\emph{IEEE Transactions on Visualization and Computer Graphics}} \bibinfo{volume}{27}, \bibinfo{number}{2} (\bibinfo{year}{2021}), \bibinfo{pages}{1860--1870}.
\newblock
\urldef\tempurl%
\url{https://doi.org/10.1109/TVCG.2020.3030340}
\showDOI{\tempurl}


\bibitem[Cukierski(2012)]%
        {titanic}
\bibfield{author}{\bibinfo{person}{Will Cukierski}.} \bibinfo{year}{2012}\natexlab{}.
\newblock \bibinfo{title}{Titanic - Machine Learning from Disaster}.
\newblock
\newblock
\urldef\tempurl%
\url{https://kaggle.com/competitions/titanic}
\showURL{%
\tempurl}


\bibitem[Cypher and Halbert(1993)]%
        {cypher1993watch}
\bibfield{author}{\bibinfo{person}{Allen Cypher} {and} \bibinfo{person}{Daniel~Conrad Halbert}.} \bibinfo{year}{1993}\natexlab{}.
\newblock \bibinfo{booktitle}{\emph{Watch what I do: programming by demonstration}}.
\newblock \bibinfo{publisher}{MIT press}.
\newblock


\bibitem[de~Souza et~al\mbox{.}(2005)]%
        {de2005A}
\bibfield{author}{\bibinfo{person}{Sergio Cozzetti~B. de Souza}, \bibinfo{person}{Nicolas Anquetil}, {and} \bibinfo{person}{K\'{a}thia~M. de Oliveira}.} \bibinfo{year}{2005}\natexlab{}.
\newblock \showarticletitle{A Study of the Documentation Essential to Software Maintenance}. In \bibinfo{booktitle}{\emph{Proceedings of the 23rd Annual International Conference on Design of Communication: Documenting \& Designing for Pervasive Information}} (Coventry, United Kingdom) \emph{(\bibinfo{series}{SIGDOC '05})}. \bibinfo{publisher}{Association for Computing Machinery}, \bibinfo{address}{New York, NY, USA}, \bibinfo{pages}{68–75}.
\newblock
\showISBNx{1595931759}
\urldef\tempurl%
\url{https://doi.org/10.1145/1085313.1085331}
\showDOI{\tempurl}


\bibitem[Dourish and Bellotti(1992)]%
        {dourish1992awareness}
\bibfield{author}{\bibinfo{person}{Paul Dourish} {and} \bibinfo{person}{Victoria Bellotti}.} \bibinfo{year}{1992}\natexlab{}.
\newblock \showarticletitle{Awareness and coordination in shared workspaces}. In \bibinfo{booktitle}{\emph{Proceedings of the 1992 ACM conference on Computer-supported cooperative work}}. \bibinfo{pages}{107--114}.
\newblock


\bibitem[Fan et~al\mbox{.}(2017)]%
        {Fan2017Balancing}
\bibfield{author}{\bibinfo{person}{Hongfei Fan}, \bibinfo{person}{Jiayao Gao}, \bibinfo{person}{Hongming Zhu}, \bibinfo{person}{Qin Liu}, \bibinfo{person}{Yang Shi}, {and} \bibinfo{person}{Chengzheng Sun}.} \bibinfo{year}{2017}\natexlab{}.
\newblock \showarticletitle{Balancing Conflict Prevention and Concurrent Work in Real-Time Collaborative Programming}. In \bibinfo{booktitle}{\emph{Proceedings of the 12th Chinese Conference on Computer Supported Cooperative Work and Social Computing}} (Chongqing, China) \emph{(\bibinfo{series}{ChineseCSCW '17})}. \bibinfo{publisher}{Association for Computing Machinery}, \bibinfo{address}{New York, NY, USA}, \bibinfo{pages}{217–220}.
\newblock
\showISBNx{9781450353526}
\urldef\tempurl%
\url{https://doi.org/10.1145/3127404.3127447}
\showDOI{\tempurl}


\bibitem[Fan et~al\mbox{.}(2012)]%
        {Fan2012ATCoPE}
\bibfield{author}{\bibinfo{person}{Hongfei Fan}, \bibinfo{person}{Chengzheng Sun}, {and} \bibinfo{person}{Haifeng Shen}.} \bibinfo{year}{2012}\natexlab{}.
\newblock \showarticletitle{ATCoPE: Any-Time Collaborative Programming Environment for Seamless Integration of Real-Time and Non-Real-Time Teamwork in Software Development}. In \bibinfo{booktitle}{\emph{Proceedings of the 2012 ACM International Conference on Supporting Group Work}} (Sanibel Island, Florida, USA) \emph{(\bibinfo{series}{GROUP '12})}. \bibinfo{publisher}{Association for Computing Machinery}, \bibinfo{address}{New York, NY, USA}, \bibinfo{pages}{107–116}.
\newblock
\showISBNx{9781450314862}
\urldef\tempurl%
\url{https://doi.org/10.1145/2389176.2389194}
\showDOI{\tempurl}


\bibitem[Fiannaca et~al\mbox{.}(2023)]%
        {Fiannaca2023Programming}
\bibfield{author}{\bibinfo{person}{Alexander~J. Fiannaca}, \bibinfo{person}{Chinmay Kulkarni}, \bibinfo{person}{Carrie~J Cai}, {and} \bibinfo{person}{Michael Terry}.} \bibinfo{year}{2023}\natexlab{}.
\newblock \showarticletitle{Programming without a Programming Language: Challenges and Opportunities for Designing Developer Tools for Prompt Programming}. In \bibinfo{booktitle}{\emph{Extended Abstracts of the 2023 CHI Conference on Human Factors in Computing Systems}} (Hamburg, Germany) \emph{(\bibinfo{series}{CHI EA '23})}. \bibinfo{publisher}{Association for Computing Machinery}, \bibinfo{address}{New York, NY, USA}, Article \bibinfo{articleno}{235}, \bibinfo{numpages}{7}~pages.
\newblock
\showISBNx{9781450394222}
\urldef\tempurl%
\url{https://doi.org/10.1145/3544549.3585737}
\showDOI{\tempurl}


\bibitem[Geiger et~al\mbox{.}(2018)]%
        {geiger2018types}
\bibfield{author}{\bibinfo{person}{R~Stuart Geiger}, \bibinfo{person}{Nelle Varoquaux}, \bibinfo{person}{Charlotte Mazel-Cabasse}, {and} \bibinfo{person}{Chris Holdgraf}.} \bibinfo{year}{2018}\natexlab{}.
\newblock \showarticletitle{The types, roles, and practices of documentation in data analytics open source software libraries: a collaborative ethnography of documentation work}.
\newblock \bibinfo{journal}{\emph{Computer Supported Cooperative Work (CSCW)}} \bibinfo{volume}{27}, \bibinfo{number}{3-6} (\bibinfo{year}{2018}), \bibinfo{pages}{767--802}.
\newblock


\bibitem[Goldman et~al\mbox{.}(2011)]%
        {Goldman2011Real}
\bibfield{author}{\bibinfo{person}{Max Goldman}, \bibinfo{person}{Greg Little}, {and} \bibinfo{person}{Robert~C. Miller}.} \bibinfo{year}{2011}\natexlab{}.
\newblock \showarticletitle{Real-Time Collaborative Coding in a Web IDE}. In \bibinfo{booktitle}{\emph{Proceedings of the 24th Annual ACM Symposium on User Interface Software and Technology}} (Santa Barbara, California, USA) \emph{(\bibinfo{series}{UIST '11})}. \bibinfo{publisher}{Association for Computing Machinery}, \bibinfo{address}{New York, NY, USA}, \bibinfo{pages}{155–164}.
\newblock
\showISBNx{9781450307161}
\urldef\tempurl%
\url{https://doi.org/10.1145/2047196.2047215}
\showDOI{\tempurl}


\bibitem[Google(2023)]%
        {firebase2023}
\bibfield{author}{\bibinfo{person}{Google}.} \bibinfo{year}{2023}\natexlab{}.
\newblock \bibinfo{title}{Firebase is an app development platform that helps you build and grow apps and games users love. Backed by Google and trusted by millions of businesses around the world.}
\newblock
\newblock
\urldef\tempurl%
\url{https://firebase.google.com/}
\showURL{%
\tempurl}


\bibitem[Green and Petre(1996)]%
        {green1996usability}
\bibfield{author}{\bibinfo{person}{Thomas R.~G. Green} {and} \bibinfo{person}{Marian Petre}.} \bibinfo{year}{1996}\natexlab{}.
\newblock \showarticletitle{Usability analysis of visual programming environments: a ‘cognitive dimensions’ framework}.
\newblock \bibinfo{journal}{\emph{Journal of Visual Languages \& Computing}} \bibinfo{volume}{7}, \bibinfo{number}{2} (\bibinfo{year}{1996}), \bibinfo{pages}{131--174}.
\newblock


\bibitem[Gutwin and Greenberg(1998)]%
        {gutwin1998effects}
\bibfield{author}{\bibinfo{person}{Carl Gutwin} {and} \bibinfo{person}{Saul Greenberg}.} \bibinfo{year}{1998}\natexlab{}.
\newblock \showarticletitle{Effects of awareness support on groupware usability}. In \bibinfo{booktitle}{\emph{Proceedings of the SIGCHI conference on Human factors in computing systems}}. \bibinfo{pages}{511--518}.
\newblock


\bibitem[Gutwin and Greenberg(2002)]%
        {Gutwin2002Descriptive}
\bibfield{author}{\bibinfo{person}{Carl Gutwin} {and} \bibinfo{person}{Saul Greenberg}.} \bibinfo{year}{2002}\natexlab{}.
\newblock \showarticletitle{A Descriptive Framework of Workspace Awareness for Real-Time Groupware}.
\newblock \bibinfo{journal}{\emph{Computer Supported Cooperative Work ({CSCW})}} \bibinfo{volume}{11}, \bibinfo{number}{3-4} (\bibinfo{date}{Sept.} \bibinfo{year}{2002}), \bibinfo{pages}{411--446}.
\newblock
\urldef\tempurl%
\url{https://doi.org/10.1023/a:1021271517844}
\showDOI{\tempurl}


\bibitem[Haverbeke(2023)]%
        {prosemirror2023}
\bibfield{author}{\bibinfo{person}{Marijn Haverbeke}.} \bibinfo{year}{2023}\natexlab{}.
\newblock \bibinfo{title}{A toolkit for building rich-text editors on the web}.
\newblock
\newblock
\urldef\tempurl%
\url{https://prosemirror.net/}
\showURL{%
\tempurl}


\bibitem[Head et~al\mbox{.}(2019)]%
        {Head2019Managing}
\bibfield{author}{\bibinfo{person}{Andrew Head}, \bibinfo{person}{Fred Hohman}, \bibinfo{person}{Titus Barik}, \bibinfo{person}{Steven~M. Drucker}, {and} \bibinfo{person}{Robert DeLine}.} \bibinfo{year}{2019}\natexlab{}.
\newblock \showarticletitle{Managing Messes in Computational Notebooks}. In \bibinfo{booktitle}{\emph{Proceedings of the 2019 CHI Conference on Human Factors in Computing Systems}} (Glasgow, Scotland Uk) \emph{(\bibinfo{series}{CHI '19})}. \bibinfo{publisher}{Association for Computing Machinery}, \bibinfo{address}{New York, NY, USA}, \bibinfo{pages}{1–12}.
\newblock
\showISBNx{9781450359702}
\urldef\tempurl%
\url{https://doi.org/10.1145/3290605.3300500}
\showDOI{\tempurl}


\bibitem[Henley and Fleming(2014)]%
        {Henley2014The}
\bibfield{author}{\bibinfo{person}{Austin~Z. Henley} {and} \bibinfo{person}{Scott~D. Fleming}.} \bibinfo{year}{2014}\natexlab{}.
\newblock \showarticletitle{The Patchworks Code Editor: Toward Faster Navigation with Less Code Arranging and Fewer Navigation Mistakes}. In \bibinfo{booktitle}{\emph{Proceedings of the SIGCHI Conference on Human Factors in Computing Systems}} (Toronto, Ontario, Canada) \emph{(\bibinfo{series}{CHI '14})}. \bibinfo{publisher}{Association for Computing Machinery}, \bibinfo{address}{New York, NY, USA}, \bibinfo{pages}{2511–2520}.
\newblock
\showISBNx{9781450324731}
\urldef\tempurl%
\url{https://doi.org/10.1145/2556288.2557073}
\showDOI{\tempurl}


\bibitem[Herbsleb et~al\mbox{.}(1995)]%
        {herbsleb1995object}
\bibfield{author}{\bibinfo{person}{James~D Herbsleb}, \bibinfo{person}{Helen Klein}, \bibinfo{person}{Gary~M Olson}, \bibinfo{person}{Hans Brunner}, \bibinfo{person}{Judith~S Olson}, {and} \bibinfo{person}{Joe Harding}.} \bibinfo{year}{1995}\natexlab{}.
\newblock \showarticletitle{Object-oriented analysis and design in software project teams}.
\newblock \bibinfo{journal}{\emph{Human--Computer Interaction}} \bibinfo{volume}{10}, \bibinfo{number}{2-3} (\bibinfo{year}{1995}), \bibinfo{pages}{249--292}.
\newblock


\bibitem[Huang et~al\mbox{.}(2023)]%
        {huang2023anpl}
\bibfield{author}{\bibinfo{person}{Di Huang}, \bibinfo{person}{Ziyuan Nan}, \bibinfo{person}{Xing Hu}, \bibinfo{person}{Pengwei Jin}, \bibinfo{person}{Shaohui Peng}, \bibinfo{person}{Yuanbo Wen}, \bibinfo{person}{Rui Zhang}, \bibinfo{person}{Zidong Du}, \bibinfo{person}{Qi Guo}, \bibinfo{person}{Yewen Pu}, {and} \bibinfo{person}{Yunji Chen}.} \bibinfo{year}{2023}\natexlab{}.
\newblock \bibinfo{title}{ANPL: Compiling Natural Programs with Interactive Decomposition}.
\newblock
\newblock
\showeprint[arxiv]{2305.18498}~[cs.PL]


\bibitem[Hutchins et~al\mbox{.}(1985)]%
        {hutchins1985direct}
\bibfield{author}{\bibinfo{person}{Edwin~L Hutchins}, \bibinfo{person}{James~D Hollan}, {and} \bibinfo{person}{Donald~A Norman}.} \bibinfo{year}{1985}\natexlab{}.
\newblock \showarticletitle{Direct manipulation interfaces}.
\newblock \bibinfo{journal}{\emph{Human--computer interaction}} \bibinfo{volume}{1}, \bibinfo{number}{4} (\bibinfo{year}{1985}), \bibinfo{pages}{311--338}.
\newblock


\bibitem[Jain et~al\mbox{.}(2022)]%
        {Jain2022Jigsaw}
\bibfield{author}{\bibinfo{person}{Naman Jain}, \bibinfo{person}{Skanda Vaidyanath}, \bibinfo{person}{Arun Iyer}, \bibinfo{person}{Nagarajan Natarajan}, \bibinfo{person}{Suresh Parthasarathy}, \bibinfo{person}{Sriram Rajamani}, {and} \bibinfo{person}{Rahul Sharma}.} \bibinfo{year}{2022}\natexlab{}.
\newblock \showarticletitle{Jigsaw: Large Language Models Meet Program Synthesis}. In \bibinfo{booktitle}{\emph{Proceedings of the 44th International Conference on Software Engineering}} (Pittsburgh, Pennsylvania) \emph{(\bibinfo{series}{ICSE '22})}. \bibinfo{publisher}{Association for Computing Machinery}, \bibinfo{address}{New York, NY, USA}, \bibinfo{pages}{1219–1231}.
\newblock
\showISBNx{9781450392211}
\urldef\tempurl%
\url{https://doi.org/10.1145/3510003.3510203}
\showDOI{\tempurl}


\bibitem[Jiang et~al\mbox{.}(2022a)]%
        {Jiang2022PromptMaker}
\bibfield{author}{\bibinfo{person}{Ellen Jiang}, \bibinfo{person}{Kristen Olson}, \bibinfo{person}{Edwin Toh}, \bibinfo{person}{Alejandra Molina}, \bibinfo{person}{Aaron Donsbach}, \bibinfo{person}{Michael Terry}, {and} \bibinfo{person}{Carrie~J Cai}.} \bibinfo{year}{2022}\natexlab{a}.
\newblock \showarticletitle{PromptMaker: Prompt-Based Prototyping with Large\&nbsp;Language\&nbsp;Models}. In \bibinfo{booktitle}{\emph{Extended Abstracts of the 2022 CHI Conference on Human Factors in Computing Systems}} (New Orleans, LA, USA) \emph{(\bibinfo{series}{CHI EA '22})}. \bibinfo{publisher}{Association for Computing Machinery}, \bibinfo{address}{New York, NY, USA}, Article \bibinfo{articleno}{35}, \bibinfo{numpages}{8}~pages.
\newblock
\showISBNx{9781450391566}
\urldef\tempurl%
\url{https://doi.org/10.1145/3491101.3503564}
\showDOI{\tempurl}


\bibitem[Jiang et~al\mbox{.}(2022b)]%
        {Jiang2022Discovering}
\bibfield{author}{\bibinfo{person}{Ellen Jiang}, \bibinfo{person}{Edwin Toh}, \bibinfo{person}{Alejandra Molina}, \bibinfo{person}{Kristen Olson}, \bibinfo{person}{Claire Kayacik}, \bibinfo{person}{Aaron Donsbach}, \bibinfo{person}{Carrie~J Cai}, {and} \bibinfo{person}{Michael Terry}.} \bibinfo{year}{2022}\natexlab{b}.
\newblock \showarticletitle{Discovering the Syntax and Strategies of Natural Language Programming with Generative Language Models}. In \bibinfo{booktitle}{\emph{Proceedings of the 2022 CHI Conference on Human Factors in Computing Systems}} (New Orleans, LA, USA) \emph{(\bibinfo{series}{CHI '22})}. \bibinfo{publisher}{Association for Computing Machinery}, \bibinfo{address}{New York, NY, USA}, Article \bibinfo{articleno}{386}, \bibinfo{numpages}{19}~pages.
\newblock
\showISBNx{9781450391573}
\urldef\tempurl%
\url{https://doi.org/10.1145/3491102.3501870}
\showDOI{\tempurl}


\bibitem[Kajko-Mattsson(2005)]%
        {kajko2005survey}
\bibfield{author}{\bibinfo{person}{Mira Kajko-Mattsson}.} \bibinfo{year}{2005}\natexlab{}.
\newblock \showarticletitle{A survey of documentation practice within corrective maintenance}.
\newblock \bibinfo{journal}{\emph{Empirical Software Engineering}}  \bibinfo{volume}{10} (\bibinfo{year}{2005}), \bibinfo{pages}{31--55}.
\newblock


\bibitem[Ko et~al\mbox{.}(2011)]%
        {Ko2011The}
\bibfield{author}{\bibinfo{person}{Amy~J. Ko}, \bibinfo{person}{Robin Abraham}, \bibinfo{person}{Laura Beckwith}, \bibinfo{person}{Alan Blackwell}, \bibinfo{person}{Margaret Burnett}, \bibinfo{person}{Martin Erwig}, \bibinfo{person}{Chris Scaffidi}, \bibinfo{person}{Joseph Lawrance}, \bibinfo{person}{Henry Lieberman}, \bibinfo{person}{Brad Myers}, \bibinfo{person}{Mary~Beth Rosson}, \bibinfo{person}{Gregg Rothermel}, \bibinfo{person}{Mary Shaw}, {and} \bibinfo{person}{Susan Wiedenbeck}.} \bibinfo{year}{2011}\natexlab{}.
\newblock \showarticletitle{The State of the Art in End-User Software Engineering}.
\newblock \bibinfo{journal}{\emph{ACM Comput. Surv.}} \bibinfo{volume}{43}, \bibinfo{number}{3}, Article \bibinfo{articleno}{21} (\bibinfo{date}{apr} \bibinfo{year}{2011}), \bibinfo{numpages}{44}~pages.
\newblock
\showISSN{0360-0300}
\urldef\tempurl%
\url{https://doi.org/10.1145/1922649.1922658}
\showDOI{\tempurl}


\bibitem[Kuznia et~al\mbox{.}(2022)]%
        {kuznia2022more}
\bibfield{author}{\bibinfo{person}{Kirby Kuznia}, \bibinfo{person}{Swaroop Mishra}, \bibinfo{person}{Mihir Parmar}, {and} \bibinfo{person}{Chitta Baral}.} \bibinfo{year}{2022}\natexlab{}.
\newblock \bibinfo{title}{Less is More: Summary of Long Instructions is Better for Program Synthesis}.
\newblock
\newblock
\showeprint[arxiv]{2203.08597}~[cs.CL]


\bibitem[Lampinen et~al\mbox{.}(2022)]%
        {lampinen2022language}
\bibfield{author}{\bibinfo{person}{Andrew~K. Lampinen}, \bibinfo{person}{Ishita Dasgupta}, \bibinfo{person}{Stephanie C.~Y. Chan}, \bibinfo{person}{Kory Matthewson}, \bibinfo{person}{Michael~Henry Tessler}, \bibinfo{person}{Antonia Creswell}, \bibinfo{person}{James~L. McClelland}, \bibinfo{person}{Jane~X. Wang}, {and} \bibinfo{person}{Felix Hill}.} \bibinfo{year}{2022}\natexlab{}.
\newblock \bibinfo{title}{Can language models learn from explanations in context?}
\newblock
\newblock
\showeprint[arxiv]{2204.02329}~[cs.CL]


\bibitem[Lauwers and Lantz(1990)]%
        {lauwers1990collaboration}
\bibfield{author}{\bibinfo{person}{J~Chris Lauwers} {and} \bibinfo{person}{Keith~A Lantz}.} \bibinfo{year}{1990}\natexlab{}.
\newblock \showarticletitle{Collaboration awareness in support of collaboration transparency: Requirements for the next generation of shared window systems}. In \bibinfo{booktitle}{\emph{Proceedings of the SIGCHI conference on Human factors in computing systems}}. \bibinfo{pages}{303--311}.
\newblock


\bibitem[Lewis et~al\mbox{.}(2013)]%
        {Lewis2013UMUX}
\bibfield{author}{\bibinfo{person}{James~R. Lewis}, \bibinfo{person}{Brian~S. Utesch}, {and} \bibinfo{person}{Deborah~E. Maher}.} \bibinfo{year}{2013}\natexlab{}.
\newblock \showarticletitle{UMUX-LITE: When There's No Time for the SUS}. In \bibinfo{booktitle}{\emph{Proceedings of the SIGCHI Conference on Human Factors in Computing Systems}} (Paris, France) \emph{(\bibinfo{series}{CHI '13})}. \bibinfo{publisher}{Association for Computing Machinery}, \bibinfo{address}{New York, NY, USA}, \bibinfo{pages}{2099–2102}.
\newblock
\showISBNx{9781450318990}
\urldef\tempurl%
\url{https://doi.org/10.1145/2470654.2481287}
\showDOI{\tempurl}


\bibitem[Li et~al\mbox{.}(2023)]%
        {Li2023CCTEST}
\bibfield{author}{\bibinfo{person}{Zongjie Li}, \bibinfo{person}{Chaozheng Wang}, \bibinfo{person}{Zhibo Liu}, \bibinfo{person}{Haoxuan Wang}, \bibinfo{person}{Dong Chen}, \bibinfo{person}{Shuai Wang}, {and} \bibinfo{person}{Cuiyun Gao}.} \bibinfo{year}{2023}\natexlab{}.
\newblock \showarticletitle{CCTEST: Testing and Repairing Code Completion Systems}. In \bibinfo{booktitle}{\emph{2023 IEEE/ACM 45th International Conference on Software Engineering (ICSE)}}. \bibinfo{pages}{1238--1250}.
\newblock
\urldef\tempurl%
\url{https://doi.org/10.1109/ICSE48619.2023.00110}
\showDOI{\tempurl}


\bibitem[Lieberman(2001)]%
        {lieberman2001your}
\bibfield{author}{\bibinfo{person}{Henry Lieberman}.} \bibinfo{year}{2001}\natexlab{}.
\newblock \bibinfo{booktitle}{\emph{Your wish is my command: Programming by example}}.
\newblock \bibinfo{publisher}{Morgan Kaufmann}.
\newblock


\bibitem[Liu and Lieberman(2005)]%
        {Liu2005Programmatic}
\bibfield{author}{\bibinfo{person}{Hugo Liu} {and} \bibinfo{person}{Henry Lieberman}.} \bibinfo{year}{2005}\natexlab{}.
\newblock \showarticletitle{Programmatic Semantics for Natural Language Interfaces}. In \bibinfo{booktitle}{\emph{CHI '05 Extended Abstracts on Human Factors in Computing Systems}} (Portland, OR, USA) \emph{(\bibinfo{series}{CHI EA '05})}. \bibinfo{publisher}{Association for Computing Machinery}, \bibinfo{address}{New York, NY, USA}, \bibinfo{pages}{1597–1600}.
\newblock
\showISBNx{1595930027}
\urldef\tempurl%
\url{https://doi.org/10.1145/1056808.1056975}
\showDOI{\tempurl}


\bibitem[Liu et~al\mbox{.}(2021b)]%
        {Liu2021To}
\bibfield{author}{\bibinfo{person}{Michael~Xieyang Liu}, \bibinfo{person}{Aniket Kittur}, {and} \bibinfo{person}{Brad~A. Myers}.} \bibinfo{year}{2021}\natexlab{b}.
\newblock \showarticletitle{To Reuse or Not To Reuse? A Framework and System for Evaluating Summarized Knowledge}.
\newblock \bibinfo{journal}{\emph{Proc. ACM Hum.-Comput. Interact.}} \bibinfo{volume}{5}, \bibinfo{number}{CSCW1}, Article \bibinfo{articleno}{166} (\bibinfo{date}{apr} \bibinfo{year}{2021}), \bibinfo{numpages}{35}~pages.
\newblock
\urldef\tempurl%
\url{https://doi.org/10.1145/3449240}
\showDOI{\tempurl}


\bibitem[Liu et~al\mbox{.}(2023a)]%
        {Liu2023What}
\bibfield{author}{\bibinfo{person}{Michael~Xieyang Liu}, \bibinfo{person}{Advait Sarkar}, \bibinfo{person}{Carina Negreanu}, \bibinfo{person}{Benjamin Zorn}, \bibinfo{person}{Jack Williams}, \bibinfo{person}{Neil Toronto}, {and} \bibinfo{person}{Andrew~D. Gordon}.} \bibinfo{year}{2023}\natexlab{a}.
\newblock \showarticletitle{“What It Wants Me To Say”: Bridging the Abstraction Gap Between End-User Programmers and Code-Generating Large Language Models}. In \bibinfo{booktitle}{\emph{Proceedings of the 2023 CHI Conference on Human Factors in Computing Systems}} (Hamburg, Germany) \emph{(\bibinfo{series}{CHI '23})}. \bibinfo{publisher}{Association for Computing Machinery}, \bibinfo{address}{New York, NY, USA}, Article \bibinfo{articleno}{598}, \bibinfo{numpages}{31}~pages.
\newblock
\showISBNx{9781450394215}
\urldef\tempurl%
\url{https://doi.org/10.1145/3544548.3580817}
\showDOI{\tempurl}


\bibitem[Liu et~al\mbox{.}(2023b)]%
        {Liu2023Pre}
\bibfield{author}{\bibinfo{person}{Pengfei Liu}, \bibinfo{person}{Weizhe Yuan}, \bibinfo{person}{Jinlan Fu}, \bibinfo{person}{Zhengbao Jiang}, \bibinfo{person}{Hiroaki Hayashi}, {and} \bibinfo{person}{Graham Neubig}.} \bibinfo{year}{2023}\natexlab{b}.
\newblock \showarticletitle{Pre-Train, Prompt, and Predict: A Systematic Survey of Prompting Methods in Natural Language Processing}.
\newblock \bibinfo{journal}{\emph{ACM Comput. Surv.}} \bibinfo{volume}{55}, \bibinfo{number}{9}, Article \bibinfo{articleno}{195} (\bibinfo{date}{jan} \bibinfo{year}{2023}), \bibinfo{numpages}{35}~pages.
\newblock
\showISSN{0360-0300}
\urldef\tempurl%
\url{https://doi.org/10.1145/3560815}
\showDOI{\tempurl}


\bibitem[Liu and Chilton(2022)]%
        {Liu2022Design}
\bibfield{author}{\bibinfo{person}{Vivian Liu} {and} \bibinfo{person}{Lydia~B Chilton}.} \bibinfo{year}{2022}\natexlab{}.
\newblock \showarticletitle{Design Guidelines for Prompt Engineering Text-to-Image Generative Models}. In \bibinfo{booktitle}{\emph{Proceedings of the 2022 CHI Conference on Human Factors in Computing Systems}} (New Orleans, LA, USA) \emph{(\bibinfo{series}{CHI '22})}. \bibinfo{publisher}{Association for Computing Machinery}, \bibinfo{address}{New York, NY, USA}, Article \bibinfo{articleno}{384}, \bibinfo{numpages}{23}~pages.
\newblock
\showISBNx{9781450391573}
\urldef\tempurl%
\url{https://doi.org/10.1145/3491102.3501825}
\showDOI{\tempurl}


\bibitem[Liu et~al\mbox{.}(2020)]%
        {Liu2020Paths}
\bibfield{author}{\bibinfo{person}{Yang Liu}, \bibinfo{person}{Tim Althoff}, {and} \bibinfo{person}{Jeffrey Heer}.} \bibinfo{year}{2020}\natexlab{}.
\newblock \showarticletitle{Paths Explored, Paths Omitted, Paths Obscured: Decision Points \& Selective Reporting in End-to-End Data Analysis}. In \bibinfo{booktitle}{\emph{Proceedings of the 2020 CHI Conference on Human Factors in Computing Systems}} (Honolulu, HI, USA) \emph{(\bibinfo{series}{CHI '20})}. \bibinfo{publisher}{Association for Computing Machinery}, \bibinfo{address}{New York, NY, USA}, \bibinfo{pages}{1–14}.
\newblock
\showISBNx{9781450367080}
\urldef\tempurl%
\url{https://doi.org/10.1145/3313831.3376533}
\showDOI{\tempurl}


\bibitem[Liu et~al\mbox{.}(2021a)]%
        {Liu2021Boba}
\bibfield{author}{\bibinfo{person}{Yang Liu}, \bibinfo{person}{Alex Kale}, \bibinfo{person}{Tim Althoff}, {and} \bibinfo{person}{Jeffrey Heer}.} \bibinfo{year}{2021}\natexlab{a}.
\newblock \showarticletitle{Boba: Authoring and Visualizing Multiverse Analyses}.
\newblock \bibinfo{journal}{\emph{IEEE Transactions on Visualization and Computer Graphics}} \bibinfo{volume}{27}, \bibinfo{number}{2} (\bibinfo{year}{2021}), \bibinfo{pages}{1753--1763}.
\newblock
\urldef\tempurl%
\url{https://doi.org/10.1109/TVCG.2020.3028985}
\showDOI{\tempurl}


\bibitem[Luger and Sellen(2016)]%
        {Luger2016Like}
\bibfield{author}{\bibinfo{person}{Ewa Luger} {and} \bibinfo{person}{Abigail Sellen}.} \bibinfo{year}{2016}\natexlab{}.
\newblock \showarticletitle{"Like Having a Really Bad PA": The Gulf between User Expectation and Experience of Conversational Agents}. In \bibinfo{booktitle}{\emph{Proceedings of the 2016 CHI Conference on Human Factors in Computing Systems}} (San Jose, California, USA) \emph{(\bibinfo{series}{CHI '16})}. \bibinfo{publisher}{Association for Computing Machinery}, \bibinfo{address}{New York, NY, USA}, \bibinfo{pages}{5286–5297}.
\newblock
\showISBNx{9781450333627}
\urldef\tempurl%
\url{https://doi.org/10.1145/2858036.2858288}
\showDOI{\tempurl}


\bibitem[Ma et~al\mbox{.}(2022)]%
        {Ma2022Integrating}
\bibfield{author}{\bibinfo{person}{Yifan Ma}, \bibinfo{person}{Batu Qi}, \bibinfo{person}{Wenhua Xu}, \bibinfo{person}{Mingjie Wang}, \bibinfo{person}{Bowen Du}, {and} \bibinfo{person}{Hongfei Fan}.} \bibinfo{year}{2022}\natexlab{}.
\newblock \showarticletitle{Integrating Real-Time and Non-Real-Time Collaborative Programming: Workflow, Techniques, and Prototypes}.
\newblock \bibinfo{journal}{\emph{Proc. ACM Hum.-Comput. Interact.}} \bibinfo{volume}{7}, \bibinfo{number}{GROUP}, Article \bibinfo{articleno}{13} (\bibinfo{date}{dec} \bibinfo{year}{2022}), \bibinfo{numpages}{19}~pages.
\newblock
\urldef\tempurl%
\url{https://doi.org/10.1145/3567563}
\showDOI{\tempurl}


\bibitem[Maalej and Robillard(2013)]%
        {Maalej2013Patterns}
\bibfield{author}{\bibinfo{person}{Walid Maalej} {and} \bibinfo{person}{Martin~P. Robillard}.} \bibinfo{year}{2013}\natexlab{}.
\newblock \showarticletitle{Patterns of Knowledge in API Reference Documentation}.
\newblock \bibinfo{journal}{\emph{IEEE Transactions on Software Engineering}} \bibinfo{volume}{39}, \bibinfo{number}{9} (\bibinfo{year}{2013}), \bibinfo{pages}{1264--1282}.
\newblock
\urldef\tempurl%
\url{https://doi.org/10.1109/TSE.2013.12}
\showDOI{\tempurl}


\bibitem[Mackeprang et~al\mbox{.}(2019)]%
        {mackeprang2019sweetSpot}
\bibfield{author}{\bibinfo{person}{Maximilian Mackeprang}, \bibinfo{person}{Claudia M\"{u}ller-Birn}, {and} \bibinfo{person}{Maximilian~Timo Stauss}.} \bibinfo{year}{2019}\natexlab{}.
\newblock \showarticletitle{Discovering the Sweet Spot of Human-Computer Configurations: A Case Study in Information Extraction}.
\newblock  \bibinfo{volume}{3}, \bibinfo{number}{CSCW}, Article \bibinfo{articleno}{195} (\bibinfo{date}{nov} \bibinfo{year}{2019}), \bibinfo{numpages}{30}~pages.
\newblock
\urldef\tempurl%
\url{https://doi.org/10.1145/3359297}
\showDOI{\tempurl}


\bibitem[Mcnutt et~al\mbox{.}(2023)]%
        {Mcnutt2023On}
\bibfield{author}{\bibinfo{person}{Andrew~M Mcnutt}, \bibinfo{person}{Chenglong Wang}, \bibinfo{person}{Robert~A Deline}, {and} \bibinfo{person}{Steven~M. Drucker}.} \bibinfo{year}{2023}\natexlab{}.
\newblock \showarticletitle{On the Design of AI-Powered Code Assistants for Notebooks}. In \bibinfo{booktitle}{\emph{Proceedings of the 2023 CHI Conference on Human Factors in Computing Systems}} (Hamburg, Germany) \emph{(\bibinfo{series}{CHI '23})}. \bibinfo{publisher}{Association for Computing Machinery}, \bibinfo{address}{New York, NY, USA}, Article \bibinfo{articleno}{434}, \bibinfo{numpages}{16}~pages.
\newblock
\showISBNx{9781450394215}
\urldef\tempurl%
\url{https://doi.org/10.1145/3544548.3580940}
\showDOI{\tempurl}


\bibitem[Microsoft(2021a)]%
        {Microsoft_2021a}
\bibfield{author}{\bibinfo{person}{Microsoft}.} \bibinfo{year}{2021}\natexlab{a}.
\newblock \bibinfo{title}{Visual studio code for the web}.
\newblock
\newblock
\urldef\tempurl%
\url{https://code.visualstudio.com/docs/editor/vscode-web}
\showURL{%
\tempurl}


\bibitem[Microsoft(2021b)]%
        {Microsoft_2021}
\bibfield{author}{\bibinfo{person}{code Microsoft}.} \bibinfo{year}{2021}\natexlab{b}.
\newblock \bibinfo{title}{Use Microsoft Live share to collaborate with Visual Studio Code}.
\newblock
\newblock
\urldef\tempurl%
\url{https://code.visualstudio.com/learn/collaboration/live-share}
\showURL{%
\tempurl}


\bibitem[Mihalcea et~al\mbox{.}(2006)]%
        {mihalcea2006nlp}
\bibfield{author}{\bibinfo{person}{Rada Mihalcea}, \bibinfo{person}{Hugo Liu}, {and} \bibinfo{person}{Henry Lieberman}.} \bibinfo{year}{2006}\natexlab{}.
\newblock \showarticletitle{NLP (natural language processing) for NLP (natural language programming)}. In \bibinfo{booktitle}{\emph{Computational Linguistics and Intelligent Text Processing: 7th International Conference, CICLing 2006, Mexico City, Mexico, February 19-25, 2006. Proceedings 7}}. Springer, \bibinfo{pages}{319--330}.
\newblock


\bibitem[Miller(1981)]%
        {Miller1981Natural}
\bibfield{author}{\bibinfo{person}{L.~A. Miller}.} \bibinfo{year}{1981}\natexlab{}.
\newblock \showarticletitle{Natural language programming: Styles, strategies, and contrasts}.
\newblock \bibinfo{journal}{\emph{IBM Systems Journal}} \bibinfo{volume}{20}, \bibinfo{number}{2} (\bibinfo{year}{1981}), \bibinfo{pages}{184--215}.
\newblock
\urldef\tempurl%
\url{https://doi.org/10.1147/sj.202.0184}
\showDOI{\tempurl}


\bibitem[Muller et~al\mbox{.}(2023)]%
        {Muller2023GenAICHI}
\bibfield{author}{\bibinfo{person}{Michael Muller}, \bibinfo{person}{Lydia~B Chilton}, \bibinfo{person}{Anna Kantosalo}, \bibinfo{person}{Q.~Vera Liao}, \bibinfo{person}{Mary~Lou Maher}, \bibinfo{person}{Charles~Patrick Martin}, {and} \bibinfo{person}{Greg Walsh}.} \bibinfo{year}{2023}\natexlab{}.
\newblock \showarticletitle{GenAICHI 2023: Generative AI and HCI at CHI 2023}. In \bibinfo{booktitle}{\emph{Extended Abstracts of the 2023 CHI Conference on Human Factors in Computing Systems}} (Hamburg, Germany) \emph{(\bibinfo{series}{CHI EA '23})}. \bibinfo{publisher}{Association for Computing Machinery}, \bibinfo{address}{New York, NY, USA}, Article \bibinfo{articleno}{350}, \bibinfo{numpages}{7}~pages.
\newblock
\showISBNx{9781450394222}
\urldef\tempurl%
\url{https://doi.org/10.1145/3544549.3573794}
\showDOI{\tempurl}


\bibitem[Muller et~al\mbox{.}(2019)]%
        {Muller2019How}
\bibfield{author}{\bibinfo{person}{Michael Muller}, \bibinfo{person}{Ingrid Lange}, \bibinfo{person}{Dakuo Wang}, \bibinfo{person}{David Piorkowski}, \bibinfo{person}{Jason Tsay}, \bibinfo{person}{Q.~Vera Liao}, \bibinfo{person}{Casey Dugan}, {and} \bibinfo{person}{Thomas Erickson}.} \bibinfo{year}{2019}\natexlab{}.
\newblock \showarticletitle{How Data Science Workers Work with Data: Discovery, Capture, Curation, Design, Creation}. In \bibinfo{booktitle}{\emph{Proceedings of the 2019 CHI Conference on Human Factors in Computing Systems}} (Glasgow, Scotland Uk) \emph{(\bibinfo{series}{CHI '19})}. \bibinfo{publisher}{Association for Computing Machinery}, \bibinfo{address}{New York, NY, USA}, \bibinfo{pages}{1–15}.
\newblock
\showISBNx{9781450359702}
\urldef\tempurl%
\url{https://doi.org/10.1145/3290605.3300356}
\showDOI{\tempurl}


\bibitem[Myers(1991)]%
        {Myers1991Separating}
\bibfield{author}{\bibinfo{person}{Brad~A. Myers}.} \bibinfo{year}{1991}\natexlab{}.
\newblock \showarticletitle{Separating Application Code from Toolkits: Eliminating the Spaghetti of Call-Backs}. In \bibinfo{booktitle}{\emph{Proceedings of the 4th Annual ACM Symposium on User Interface Software and Technology}} (Hilton Head, South Carolina, USA) \emph{(\bibinfo{series}{UIST '91})}. \bibinfo{publisher}{Association for Computing Machinery}, \bibinfo{address}{New York, NY, USA}, \bibinfo{pages}{211–220}.
\newblock
\showISBNx{0897914511}
\urldef\tempurl%
\url{https://doi.org/10.1145/120782.120805}
\showDOI{\tempurl}


\bibitem[Oney et~al\mbox{.}(2018)]%
        {Oney2018Creating}
\bibfield{author}{\bibinfo{person}{Steve Oney}, \bibinfo{person}{Christopher Brooks}, {and} \bibinfo{person}{Paul Resnick}.} \bibinfo{year}{2018}\natexlab{}.
\newblock \showarticletitle{Creating Guided Code Explanations with Chat.Codes}.
\newblock \bibinfo{journal}{\emph{Proc. ACM Hum.-Comput. Interact.}} \bibinfo{volume}{2}, \bibinfo{number}{CSCW}, Article \bibinfo{articleno}{131} (\bibinfo{date}{nov} \bibinfo{year}{2018}), \bibinfo{numpages}{20}~pages.
\newblock
\urldef\tempurl%
\url{https://doi.org/10.1145/3274400}
\showDOI{\tempurl}


\bibitem[OpenAI(2023)]%
        {openai2023gpt4}
\bibfield{author}{\bibinfo{person}{OpenAI}.} \bibinfo{year}{2023}\natexlab{}.
\newblock \bibinfo{title}{GPT-4 Technical Report}.
\newblock
\newblock
\showeprint[arxiv]{2303.08774}~[cs.CL]


\bibitem[Padioleau et~al\mbox{.}(2009)]%
        {Padioleau2009Listening}
\bibfield{author}{\bibinfo{person}{Yoann Padioleau}, \bibinfo{person}{Lin Tan}, {and} \bibinfo{person}{Yuanyuan Zhou}.} \bibinfo{year}{2009}\natexlab{}.
\newblock \showarticletitle{Listening to programmers — Taxonomies and characteristics of comments in operating system code}. In \bibinfo{booktitle}{\emph{2009 IEEE 31st International Conference on Software Engineering}}. \bibinfo{pages}{331--341}.
\newblock
\urldef\tempurl%
\url{https://doi.org/10.1109/ICSE.2009.5070533}
\showDOI{\tempurl}


\bibitem[Pang et~al\mbox{.}(2022a)]%
        {pang2022data}
\bibfield{author}{\bibinfo{person}{Rock~Yuren Pang}, \bibinfo{person}{Ruotong Wang}, \bibinfo{person}{Joely Nelson}, {and} \bibinfo{person}{Leilani Battle}.} \bibinfo{year}{2022}\natexlab{a}.
\newblock \showarticletitle{How Do Data Science Workers Communicate Intermediate Results?}. In \bibinfo{booktitle}{\emph{2022 IEEE Visualization in Data Science (VDS)}}. IEEE, \bibinfo{pages}{46--54}.
\newblock


\bibitem[Pang et~al\mbox{.}(2022b)]%
        {Pang2022How}
\bibfield{author}{\bibinfo{person}{Rock~Yuren Pang}, \bibinfo{person}{Ruotong Wang}, \bibinfo{person}{Joely Nelson}, {and} \bibinfo{person}{Leilani Battle}.} \bibinfo{year}{2022}\natexlab{b}.
\newblock \showarticletitle{How Do Data Science Workers Communicate Intermediate Results?}. In \bibinfo{booktitle}{\emph{2022 IEEE Visualization in Data Science (VDS)}}. \bibinfo{pages}{46--54}.
\newblock
\urldef\tempurl%
\url{https://doi.org/10.1109/VDS57266.2022.00010}
\showDOI{\tempurl}


\bibitem[Park et~al\mbox{.}(2018)]%
        {Park2018Post}
\bibfield{author}{\bibinfo{person}{Soya Park}, \bibinfo{person}{Amy~X. Zhang}, {and} \bibinfo{person}{David~R. Karger}.} \bibinfo{year}{2018}\natexlab{}.
\newblock \showarticletitle{Post-Literate Programming: Linking Discussion and Code in Software Development Teams}. In \bibinfo{booktitle}{\emph{Adjunct Proceedings of the 31st Annual ACM Symposium on User Interface Software and Technology}} (Berlin, Germany) \emph{(\bibinfo{series}{UIST '18 Adjunct})}. \bibinfo{publisher}{Association for Computing Machinery}, \bibinfo{address}{New York, NY, USA}, \bibinfo{pages}{51–53}.
\newblock
\showISBNx{9781450359498}
\urldef\tempurl%
\url{https://doi.org/10.1145/3266037.3266098}
\showDOI{\tempurl}


\bibitem[Pearce et~al\mbox{.}(2023)]%
        {Pearce2023Examining}
\bibfield{author}{\bibinfo{person}{Hammond Pearce}, \bibinfo{person}{Benjamin Tan}, \bibinfo{person}{Baleegh Ahmad}, \bibinfo{person}{Ramesh Karri}, {and} \bibinfo{person}{Brendan Dolan-Gavitt}.} \bibinfo{year}{2023}\natexlab{}.
\newblock \showarticletitle{Examining Zero-Shot Vulnerability Repair with Large Language Models}. In \bibinfo{booktitle}{\emph{2023 IEEE Symposium on Security and Privacy (SP)}}. \bibinfo{pages}{2339--2356}.
\newblock
\urldef\tempurl%
\url{https://doi.org/10.1109/SP46215.2023.10179324}
\showDOI{\tempurl}


\bibitem[Posner and Baecker(1992)]%
        {Posner1992How}
\bibfield{author}{\bibinfo{person}{I.R. Posner} {and} \bibinfo{person}{R.M. Baecker}.} \bibinfo{year}{1992}\natexlab{}.
\newblock \showarticletitle{How people write together (groupware)}. In \bibinfo{booktitle}{\emph{Proceedings of the Twenty-Fifth Hawaii International Conference on System Sciences}}, Vol.~\bibinfo{volume}{iv}. \bibinfo{pages}{127--138 vol.4}.
\newblock
\urldef\tempurl%
\url{https://doi.org/10.1109/HICSS.1992.183420}
\showDOI{\tempurl}


\bibitem[Pyodide(2023)]%
        {pyodide2023}
\bibfield{author}{\bibinfo{person}{Pyodide}.} \bibinfo{year}{2023}\natexlab{}.
\newblock \bibinfo{title}{Pyodide is a Python distribution for the browser and Node.js based on WebAssembly.}
\newblock
\newblock
\urldef\tempurl%
\url{https://github.com/pyodide/pyodide}
\showURL{%
\tempurl}


\bibitem[Quaranta et~al\mbox{.}(2022)]%
        {Quaranta2022Eliciting}
\bibfield{author}{\bibinfo{person}{Luigi Quaranta}, \bibinfo{person}{Fabio Calefato}, {and} \bibinfo{person}{Filippo Lanubile}.} \bibinfo{year}{2022}\natexlab{}.
\newblock \showarticletitle{Eliciting Best Practices for Collaboration with Computational Notebooks}.
\newblock \bibinfo{journal}{\emph{Proc. ACM Hum.-Comput. Interact.}} \bibinfo{volume}{6}, \bibinfo{number}{CSCW1}, Article \bibinfo{articleno}{87} (\bibinfo{date}{apr} \bibinfo{year}{2022}), \bibinfo{numpages}{41}~pages.
\newblock
\urldef\tempurl%
\url{https://doi.org/10.1145/3512934}
\showDOI{\tempurl}


\bibitem[Reynolds and McDonell(2021)]%
        {Reynolds2021Prompt}
\bibfield{author}{\bibinfo{person}{Laria Reynolds} {and} \bibinfo{person}{Kyle McDonell}.} \bibinfo{year}{2021}\natexlab{}.
\newblock \showarticletitle{Prompt Programming for Large Language Models: Beyond the Few-Shot Paradigm}. In \bibinfo{booktitle}{\emph{Extended Abstracts of the 2021 CHI Conference on Human Factors in Computing Systems}} (Yokohama, Japan) \emph{(\bibinfo{series}{CHI EA '21})}. \bibinfo{publisher}{Association for Computing Machinery}, \bibinfo{address}{New York, NY, USA}, Article \bibinfo{articleno}{314}, \bibinfo{numpages}{7}~pages.
\newblock
\showISBNx{9781450380959}
\urldef\tempurl%
\url{https://doi.org/10.1145/3411763.3451760}
\showDOI{\tempurl}


\bibitem[Ritschel et~al\mbox{.}(2022)]%
        {Ritschel2022Can}
\bibfield{author}{\bibinfo{person}{Nico Ritschel}, \bibinfo{person}{Felipe Fronchetti}, \bibinfo{person}{Reid Holmes}, \bibinfo{person}{Ronald Garcia}, {and} \bibinfo{person}{David~C. Shepherd}.} \bibinfo{year}{2022}\natexlab{}.
\newblock \showarticletitle{Can Guided Decomposition Help End-Users Write Larger Block-Based Programs? A Mobile Robot Experiment}.
\newblock \bibinfo{journal}{\emph{Proc. ACM Program. Lang.}} \bibinfo{volume}{6}, \bibinfo{number}{OOPSLA2}, Article \bibinfo{articleno}{133} (\bibinfo{date}{oct} \bibinfo{year}{2022}), \bibinfo{numpages}{26}~pages.
\newblock
\urldef\tempurl%
\url{https://doi.org/10.1145/3563296}
\showDOI{\tempurl}


\bibitem[Roehm et~al\mbox{.}(2012)]%
        {Roehm2012How}
\bibfield{author}{\bibinfo{person}{Tobias Roehm}, \bibinfo{person}{Rebecca Tiarks}, \bibinfo{person}{Rainer Koschke}, {and} \bibinfo{person}{Walid Maalej}.} \bibinfo{year}{2012}\natexlab{}.
\newblock \showarticletitle{How do professional developers comprehend software?}. In \bibinfo{booktitle}{\emph{2012 34th International Conference on Software Engineering (ICSE)}}. \bibinfo{pages}{255--265}.
\newblock
\urldef\tempurl%
\url{https://doi.org/10.1109/ICSE.2012.6227188}
\showDOI{\tempurl}


\bibitem[Ross et~al\mbox{.}(2023)]%
        {ross2023programmer}
\bibfield{author}{\bibinfo{person}{Steven~I Ross}, \bibinfo{person}{Fernando Martinez}, \bibinfo{person}{Stephanie Houde}, \bibinfo{person}{Michael Muller}, {and} \bibinfo{person}{Justin~D Weisz}.} \bibinfo{year}{2023}\natexlab{}.
\newblock \showarticletitle{The programmer’s assistant: Conversational interaction with a large language model for software development}. In \bibinfo{booktitle}{\emph{Proceedings of the 28th International Conference on Intelligent User Interfaces}}. \bibinfo{pages}{491--514}.
\newblock


\bibitem[Sarkar et~al\mbox{.}(2022)]%
        {sarkar2022like}
\bibfield{author}{\bibinfo{person}{Advait Sarkar}, \bibinfo{person}{Andrew~D. Gordon}, \bibinfo{person}{Carina Negreanu}, \bibinfo{person}{Christian Poelitz}, \bibinfo{person}{Sruti~Srinivasa Ragavan}, {and} \bibinfo{person}{Ben Zorn}.} \bibinfo{year}{2022}\natexlab{}.
\newblock \bibinfo{title}{What is it like to program with artificial intelligence?}
\newblock
\newblock
\urldef\tempurl%
\url{https://doi.org/10.48550/arXiv.2208.06213}
\showDOI{\tempurl}
\showeprint[arxiv]{2208.06213}~[cs.HC]


\bibitem[Sarma et~al\mbox{.}(2023)]%
        {Sarma2023Multiverse}
\bibfield{author}{\bibinfo{person}{Abhraneel Sarma}, \bibinfo{person}{Alex Kale}, \bibinfo{person}{Michael~Jongho Moon}, \bibinfo{person}{Nathan Taback}, \bibinfo{person}{Fanny Chevalier}, \bibinfo{person}{Jessica Hullman}, {and} \bibinfo{person}{Matthew Kay}.} \bibinfo{year}{2023}\natexlab{}.
\newblock \showarticletitle{Multiverse: Multiplexing Alternative Data Analyses in R Notebooks}. In \bibinfo{booktitle}{\emph{Proceedings of the 2023 CHI Conference on Human Factors in Computing Systems}} (Hamburg, Germany) \emph{(\bibinfo{series}{CHI '23})}. \bibinfo{publisher}{Association for Computing Machinery}, \bibinfo{address}{New York, NY, USA}, Article \bibinfo{articleno}{148}, \bibinfo{numpages}{15}~pages.
\newblock
\showISBNx{9781450394215}
\urldef\tempurl%
\url{https://doi.org/10.1145/3544548.3580726}
\showDOI{\tempurl}


\bibitem[Shi et~al\mbox{.}(2011)]%
        {shi2011empirical}
\bibfield{author}{\bibinfo{person}{Lin Shi}, \bibinfo{person}{Hao Zhong}, \bibinfo{person}{Tao Xie}, {and} \bibinfo{person}{Mingshu Li}.} \bibinfo{year}{2011}\natexlab{}.
\newblock \showarticletitle{An empirical study on evolution of API documentation}. In \bibinfo{booktitle}{\emph{Fundamental Approaches to Software Engineering: 14th International Conference, FASE 2011, Held as Part of the Joint European Conferences on Theory and Practice of Software, ETAPS 2011, Saarbr{\"u}cken, Germany, March 26--April 3, 2011. Proceedings 14}}. Springer, \bibinfo{pages}{416--431}.
\newblock


\bibitem[Strobelt et~al\mbox{.}(2023)]%
        {strobelt2023interactive}
\bibfield{author}{\bibinfo{person}{Hendrik Strobelt}, \bibinfo{person}{Albert Webson}, \bibinfo{person}{Victor Sanh}, \bibinfo{person}{Benjamin Hoover}, \bibinfo{person}{Johanna Beyer}, \bibinfo{person}{Hanspeter Pfister}, {and} \bibinfo{person}{Alexander~M. Rush}.} \bibinfo{year}{2023}\natexlab{}.
\newblock \showarticletitle{Interactive and Visual Prompt Engineering for Ad-hoc Task Adaptation with Large Language Models}.
\newblock \bibinfo{journal}{\emph{IEEE Transactions on Visualization and Computer Graphics}} \bibinfo{volume}{29}, \bibinfo{number}{1} (\bibinfo{year}{2023}), \bibinfo{pages}{1146--1156}.
\newblock
\urldef\tempurl%
\url{https://doi.org/10.1109/TVCG.2022.3209479}
\showDOI{\tempurl}


\bibitem[Tiptap(2023)]%
        {tiptap2023}
\bibfield{author}{\bibinfo{person}{Tiptap}.} \bibinfo{year}{2023}\natexlab{}.
\newblock \bibinfo{title}{Tiptap is a suite of open source content editing and real-time collaboration tools for developers building apps like Notion or Google Docs.}
\newblock
\newblock
\urldef\tempurl%
\url{https://tiptap.dev/}
\showURL{%
\tempurl}


\bibitem[Trummer(2022)]%
        {trummer2022codexdb}
\bibfield{author}{\bibinfo{person}{Immanuel Trummer}.} \bibinfo{year}{2022}\natexlab{}.
\newblock \bibinfo{title}{CodexDB: Generating Code for Processing SQL Queries using GPT-3 Codex}.
\newblock
\newblock
\showeprint[arxiv]{2204.08941}~[cs.DB]


\bibitem[Vaithilingam et~al\mbox{.}(2023)]%
        {Vaithilingam2023Towards}
\bibfield{author}{\bibinfo{person}{Priyan Vaithilingam}, \bibinfo{person}{Elena~L. Glassman}, \bibinfo{person}{Peter Groenwegen}, \bibinfo{person}{Sumit Gulwani}, \bibinfo{person}{Austin~Z. Henley}, \bibinfo{person}{Rohan Malpani}, \bibinfo{person}{David Pugh}, \bibinfo{person}{Arjun Radhakrishna}, \bibinfo{person}{Gustavo Soares}, \bibinfo{person}{Joey Wang}, {and} \bibinfo{person}{Aaron Yim}.} \bibinfo{year}{2023}\natexlab{}.
\newblock \showarticletitle{Towards More Effective AI-Assisted Programming: A Systematic Design Exploration to Improve Visual Studio IntelliCode’s User Experience}. In \bibinfo{booktitle}{\emph{2023 IEEE/ACM 45th International Conference on Software Engineering: Software Engineering in Practice (ICSE-SEIP)}}. \bibinfo{pages}{185--195}.
\newblock
\urldef\tempurl%
\url{https://doi.org/10.1109/ICSE-SEIP58684.2023.00022}
\showDOI{\tempurl}


\bibitem[Vasconcelos et~al\mbox{.}(2023)]%
        {vasconcelos2023generation}
\bibfield{author}{\bibinfo{person}{Helena Vasconcelos}, \bibinfo{person}{Gagan Bansal}, \bibinfo{person}{Adam Fourney}, \bibinfo{person}{Q~Vera Liao}, {and} \bibinfo{person}{Jennifer~Wortman Vaughan}.} \bibinfo{year}{2023}\natexlab{}.
\newblock \showarticletitle{Generation probabilities are not enough: Exploring the effectiveness of uncertainty highlighting in AI-powered code completions}.
\newblock \bibinfo{journal}{\emph{arXiv preprint arXiv:2302.07248}} (\bibinfo{year}{2023}).
\newblock


\bibitem[Wang(2022)]%
        {Wang2022Improving}
\bibfield{author}{\bibinfo{person}{April~Yi Wang}.} \bibinfo{year}{2022}\natexlab{}.
\newblock \showarticletitle{Improving Real-Time Collaborative Data Science Through Context-Aware Mechanisms}. In \bibinfo{booktitle}{\emph{2022 IEEE Symposium on Visual Languages and Human-Centric Computing (VL/HCC)}}. \bibinfo{pages}{1--3}.
\newblock
\urldef\tempurl%
\url{https://doi.org/10.1109/VL/HCC53370.2022.9833140}
\showDOI{\tempurl}


\bibitem[Wang et~al\mbox{.}(2019a)]%
        {Wang2019How}
\bibfield{author}{\bibinfo{person}{April~Yi Wang}, \bibinfo{person}{Anant Mittal}, \bibinfo{person}{Christopher Brooks}, {and} \bibinfo{person}{Steve Oney}.} \bibinfo{year}{2019}\natexlab{a}.
\newblock \showarticletitle{How Data Scientists Use Computational Notebooks for Real-Time Collaboration}.
\newblock \bibinfo{journal}{\emph{Proc. ACM Hum.-Comput. Interact.}} \bibinfo{volume}{3}, \bibinfo{number}{CSCW}, Article \bibinfo{articleno}{39} (\bibinfo{date}{nov} \bibinfo{year}{2019}), \bibinfo{numpages}{30}~pages.
\newblock
\urldef\tempurl%
\url{https://doi.org/10.1145/3359141}
\showDOI{\tempurl}


\bibitem[Wang et~al\mbox{.}(2022)]%
        {Wang2022Documentation}
\bibfield{author}{\bibinfo{person}{April~Yi Wang}, \bibinfo{person}{Dakuo Wang}, \bibinfo{person}{Jaimie Drozdal}, \bibinfo{person}{Michael Muller}, \bibinfo{person}{Soya Park}, \bibinfo{person}{Justin~D. Weisz}, \bibinfo{person}{Xuye Liu}, \bibinfo{person}{Lingfei Wu}, {and} \bibinfo{person}{Casey Dugan}.} \bibinfo{year}{2022}\natexlab{}.
\newblock \showarticletitle{Documentation Matters: Human-Centered AI System to Assist Data Science Code Documentation in Computational Notebooks}.
\newblock \bibinfo{journal}{\emph{ACM Trans. Comput.-Hum. Interact.}} \bibinfo{volume}{29}, \bibinfo{number}{2}, Article \bibinfo{articleno}{17} (\bibinfo{date}{jan} \bibinfo{year}{2022}), \bibinfo{numpages}{33}~pages.
\newblock
\showISSN{1073-0516}
\urldef\tempurl%
\url{https://doi.org/10.1145/3489465}
\showDOI{\tempurl}


\bibitem[Wang et~al\mbox{.}(2020b)]%
        {Wang2020Callisto}
\bibfield{author}{\bibinfo{person}{April~Yi Wang}, \bibinfo{person}{Zihan Wu}, \bibinfo{person}{Christopher Brooks}, {and} \bibinfo{person}{Steve Oney}.} \bibinfo{year}{2020}\natexlab{b}.
\newblock \showarticletitle{Callisto: Capturing the "Why" by Connecting Conversations with Computational Narratives}. In \bibinfo{booktitle}{\emph{Proceedings of the 2020 CHI Conference on Human Factors in Computing Systems}} (Honolulu, HI, USA) \emph{(\bibinfo{series}{CHI '20})}. \bibinfo{publisher}{Association for Computing Machinery}, \bibinfo{address}{New York, NY, USA}, \bibinfo{pages}{1–13}.
\newblock
\showISBNx{9781450367080}
\urldef\tempurl%
\url{https://doi.org/10.1145/3313831.3376740}
\showDOI{\tempurl}


\bibitem[Wang et~al\mbox{.}(2020a)]%
        {Wang2020From}
\bibfield{author}{\bibinfo{person}{Dakuo Wang}, \bibinfo{person}{Elizabeth Churchill}, \bibinfo{person}{Pattie Maes}, \bibinfo{person}{Xiangmin Fan}, \bibinfo{person}{Ben Shneiderman}, \bibinfo{person}{Yuanchun Shi}, {and} \bibinfo{person}{Qianying Wang}.} \bibinfo{year}{2020}\natexlab{a}.
\newblock \showarticletitle{From Human-Human Collaboration to Human-AI Collaboration: Designing AI Systems That Can Work Together with People}. In \bibinfo{booktitle}{\emph{Extended Abstracts of the 2020 CHI Conference on Human Factors in Computing Systems}} (Honolulu, HI, USA) \emph{(\bibinfo{series}{CHI EA '20})}. \bibinfo{publisher}{Association for Computing Machinery}, \bibinfo{address}{New York, NY, USA}, \bibinfo{pages}{1–6}.
\newblock
\showISBNx{9781450368193}
\urldef\tempurl%
\url{https://doi.org/10.1145/3334480.3381069}
\showDOI{\tempurl}


\bibitem[Wang et~al\mbox{.}(2019b)]%
        {Wang2019Human}
\bibfield{author}{\bibinfo{person}{Dakuo Wang}, \bibinfo{person}{Justin~D. Weisz}, \bibinfo{person}{Michael Muller}, \bibinfo{person}{Parikshit Ram}, \bibinfo{person}{Werner Geyer}, \bibinfo{person}{Casey Dugan}, \bibinfo{person}{Yla Tausczik}, \bibinfo{person}{Horst Samulowitz}, {and} \bibinfo{person}{Alexander Gray}.} \bibinfo{year}{2019}\natexlab{b}.
\newblock \showarticletitle{Human-AI Collaboration in Data Science: Exploring Data Scientists' Perceptions of Automated AI}.
\newblock \bibinfo{journal}{\emph{Proc. ACM Hum.-Comput. Interact.}} \bibinfo{volume}{3}, \bibinfo{number}{CSCW}, Article \bibinfo{articleno}{211} (\bibinfo{date}{nov} \bibinfo{year}{2019}), \bibinfo{numpages}{24}~pages.
\newblock
\urldef\tempurl%
\url{https://doi.org/10.1145/3359313}
\showDOI{\tempurl}


\bibitem[Wang et~al\mbox{.}(2023)]%
        {wang2023reelframe1}
\bibfield{author}{\bibinfo{person}{Sitong Wang}, \bibinfo{person}{Samia Menon}, \bibinfo{person}{Tao Long}, \bibinfo{person}{Keren Henderson}, \bibinfo{person}{Dingzeyu Li}, \bibinfo{person}{Kevin Crowston}, \bibinfo{person}{Mark Hansen}, \bibinfo{person}{Jeffrey~V Nickerson}, {and} \bibinfo{person}{Lydia~B Chilton}.} \bibinfo{year}{2023}\natexlab{}.
\newblock \showarticletitle{ReelFramer: Co-creating News Reels on Social Media with Generative AI}.
\newblock \bibinfo{journal}{\emph{arXiv preprint arXiv:2304.09653}} (\bibinfo{year}{2023}).
\newblock


\bibitem[Wei et~al\mbox{.}(2023)]%
        {wei2023chainofthought}
\bibfield{author}{\bibinfo{person}{Jason Wei}, \bibinfo{person}{Xuezhi Wang}, \bibinfo{person}{Dale Schuurmans}, \bibinfo{person}{Maarten Bosma}, \bibinfo{person}{Brian Ichter}, \bibinfo{person}{Fei Xia}, \bibinfo{person}{Ed Chi}, \bibinfo{person}{Quoc Le}, {and} \bibinfo{person}{Denny Zhou}.} \bibinfo{year}{2023}\natexlab{}.
\newblock \bibinfo{title}{Chain-of-Thought Prompting Elicits Reasoning in Large Language Models}.
\newblock
\newblock
\showeprint[arxiv]{2201.11903}~[cs.CL]


\bibitem[Wenskovitch et~al\mbox{.}(2019)]%
        {Wenskovitch2019Albireo}
\bibfield{author}{\bibinfo{person}{John Wenskovitch}, \bibinfo{person}{Jian Zhao}, \bibinfo{person}{Scott Carter}, \bibinfo{person}{Matthew Cooper}, {and} \bibinfo{person}{Chris North}.} \bibinfo{year}{2019}\natexlab{}.
\newblock \showarticletitle{Albireo: An Interactive Tool for Visually Summarizing Computational Notebook Structure}. In \bibinfo{booktitle}{\emph{2019 IEEE Visualization in Data Science (VDS)}}. \bibinfo{pages}{1--10}.
\newblock
\urldef\tempurl%
\url{https://doi.org/10.1109/VDS48975.2019.8973385}
\showDOI{\tempurl}


\bibitem[Williams et~al\mbox{.}(2000)]%
        {Williams2000Strengthening}
\bibfield{author}{\bibinfo{person}{L. Williams}, \bibinfo{person}{R.R. Kessler}, \bibinfo{person}{W. Cunningham}, {and} \bibinfo{person}{R. Jeffries}.} \bibinfo{year}{2000}\natexlab{}.
\newblock \showarticletitle{Strengthening the case for pair programming}.
\newblock \bibinfo{journal}{\emph{IEEE Software}} \bibinfo{volume}{17}, \bibinfo{number}{4} (\bibinfo{year}{2000}), \bibinfo{pages}{19--25}.
\newblock
\urldef\tempurl%
\url{https://doi.org/10.1109/52.854064}
\showDOI{\tempurl}


\bibitem[Wu et~al\mbox{.}(2022)]%
        {Wu2022AI}
\bibfield{author}{\bibinfo{person}{Tongshuang Wu}, \bibinfo{person}{Michael Terry}, {and} \bibinfo{person}{Carrie~Jun Cai}.} \bibinfo{year}{2022}\natexlab{}.
\newblock \showarticletitle{AI Chains: Transparent and Controllable Human-AI Interaction by Chaining Large Language Model Prompts}. In \bibinfo{booktitle}{\emph{Proceedings of the 2022 CHI Conference on Human Factors in Computing Systems}} (New Orleans, LA, USA) \emph{(\bibinfo{series}{CHI '22})}. \bibinfo{publisher}{Association for Computing Machinery}, \bibinfo{address}{New York, NY, USA}, Article \bibinfo{articleno}{385}, \bibinfo{numpages}{22}~pages.
\newblock
\showISBNx{9781450391573}
\urldef\tempurl%
\url{https://doi.org/10.1145/3491102.3517582}
\showDOI{\tempurl}


\bibitem[Yen et~al\mbox{.}(2023)]%
        {yen2023coladder}
\bibfield{author}{\bibinfo{person}{Ryan Yen}, \bibinfo{person}{Jiawen Zhu}, \bibinfo{person}{Sangho Suh}, \bibinfo{person}{Haijun Xia}, {and} \bibinfo{person}{Jian Zhao}.} \bibinfo{year}{2023}\natexlab{}.
\newblock \showarticletitle{CoLadder: Supporting Programmers with Hierarchical Code Generation in Multi-Level Abstraction}.
\newblock \bibinfo{journal}{\emph{arXiv preprint arXiv:2310.08699}} (\bibinfo{year}{2023}).
\newblock


\bibitem[Ying and Boyer(2020)]%
        {Ying2020Understanding}
\bibfield{author}{\bibinfo{person}{Kimberly~Michelle Ying} {and} \bibinfo{person}{Kristy~Elizabeth Boyer}.} \bibinfo{year}{2020}\natexlab{}.
\newblock \showarticletitle{Understanding Students' Needs for Better Collaborative Coding Tools}. In \bibinfo{booktitle}{\emph{Extended Abstracts of the 2020 CHI Conference on Human Factors in Computing Systems}} (<conf-loc>, <city>Honolulu</city>, <state>HI</state>, <country>USA</country>, </conf-loc>) \emph{(\bibinfo{series}{CHI EA '20})}. \bibinfo{publisher}{Association for Computing Machinery}, \bibinfo{address}{New York, NY, USA}, \bibinfo{pages}{1–8}.
\newblock
\showISBNx{9781450368193}
\urldef\tempurl%
\url{https://doi.org/10.1145/3334480.3383068}
\showDOI{\tempurl}


\bibitem[Zhang et~al\mbox{.}(2023)]%
        {Zhang2023VISAR}
\bibfield{author}{\bibinfo{person}{Zheng Zhang}, \bibinfo{person}{Jie Gao}, \bibinfo{person}{Ranjodh~Singh Dhaliwal}, {and} \bibinfo{person}{Toby Jia-Jun Li}.} \bibinfo{year}{2023}\natexlab{}.
\newblock \showarticletitle{VISAR: A Human-AI Argumentative Writing Assistant with Visual Programming and Rapid Draft Prototyping}. In \bibinfo{booktitle}{\emph{Proceedings of the 36th Annual ACM Symposium on User Interface Software and Technology}} (San Francisco, CA, USA) \emph{(\bibinfo{series}{UIST '23})}. \bibinfo{publisher}{Association for Computing Machinery}, \bibinfo{address}{New York, NY, USA}, Article \bibinfo{articleno}{5}, \bibinfo{numpages}{30}~pages.
\newblock
\showISBNx{9798400701320}
\urldef\tempurl%
\url{https://doi.org/10.1145/3586183.3606800}
\showDOI{\tempurl}


\end{thebibliography}

\appendix
\rv{
\section{System Design}
\subsection{System Architecture}
\begin{figure}[H]
    \centering
    \includegraphics[width=0.95\linewidth]{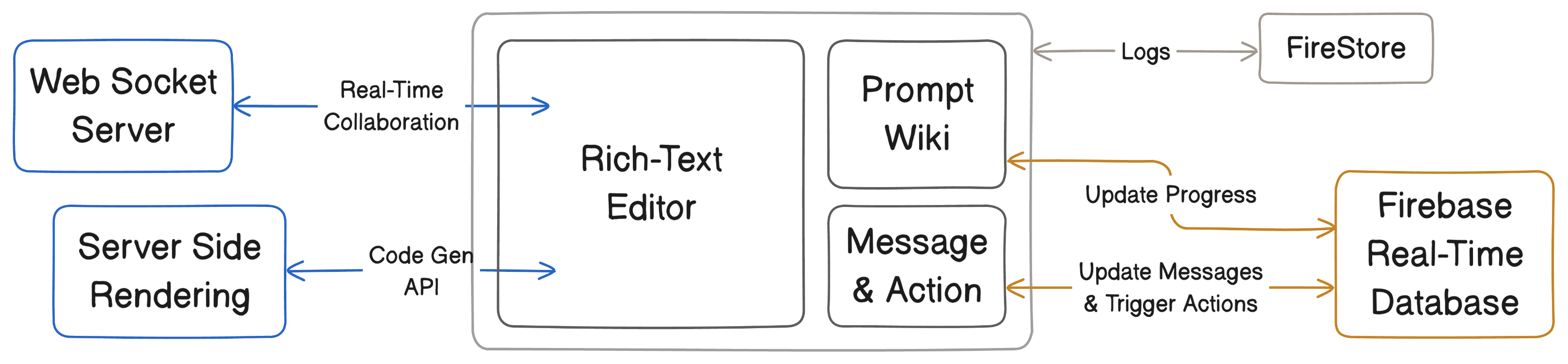}
    \caption{The system architecture of \sys{} consists of a web socket server and a server-side rendering route for invoking the generative AI API. All messages and triggered actions are managed and updated using the Firebase Real-Time Database, while logs are stored in Firestore.}
    \label{fig:sys-architecture}
\end{figure}

\subsection{Prompt Templates}
\label{appendix:prompt}
\begin{figure}[H]
    \centering
    \includegraphics[width=1\linewidth]{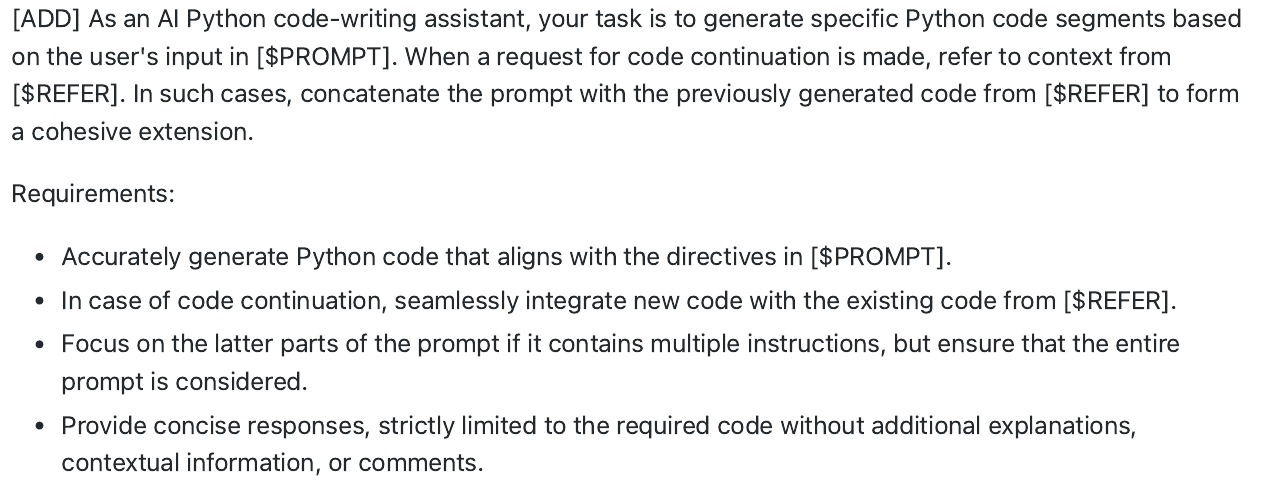}
    \caption{Prompt template for Adding prompt blocks.}
    \label{fig:prompt-add}
\end{figure}

\begin{figure}[H]
    \centering
    \includegraphics[width=1\linewidth]{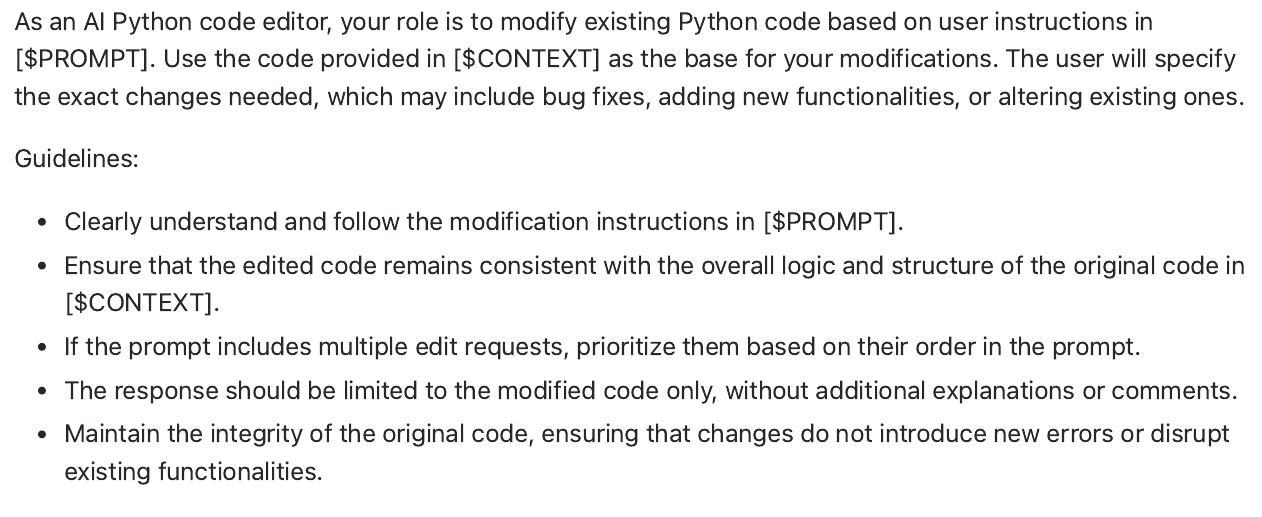}
    \caption{Prompt template for editing existing prompt blocks.}
    \label{fig:prompt-edit}
\end{figure}

\begin{figure}[H]
    \centering
    \includegraphics[width=1\linewidth]{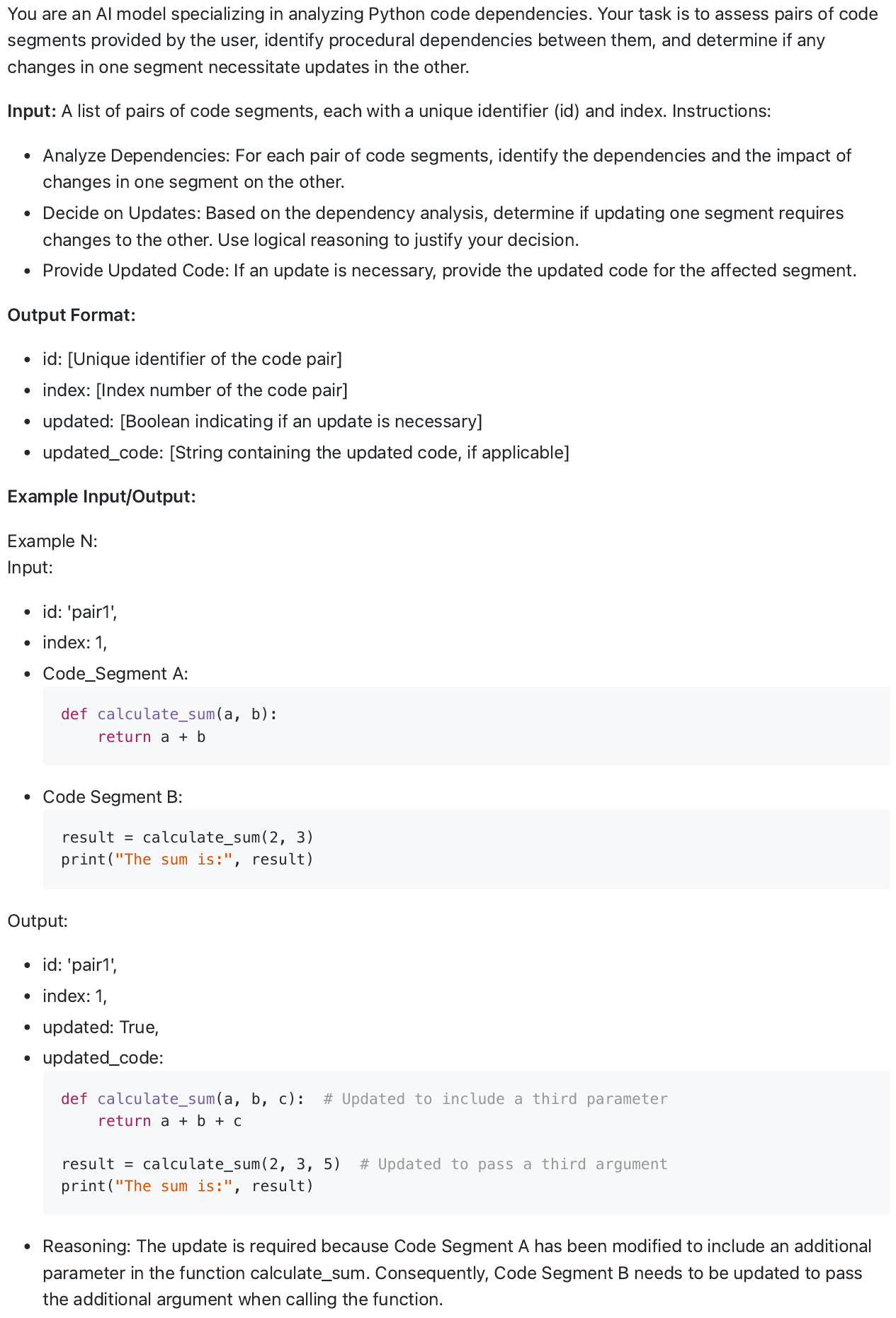}
    \caption{Prompt template for \reciprocatelink{} mechanism.}
    \label{fig:prompt-link}
\end{figure}

\begin{figure}[H]
    \centering
    \includegraphics[width=1\linewidth]{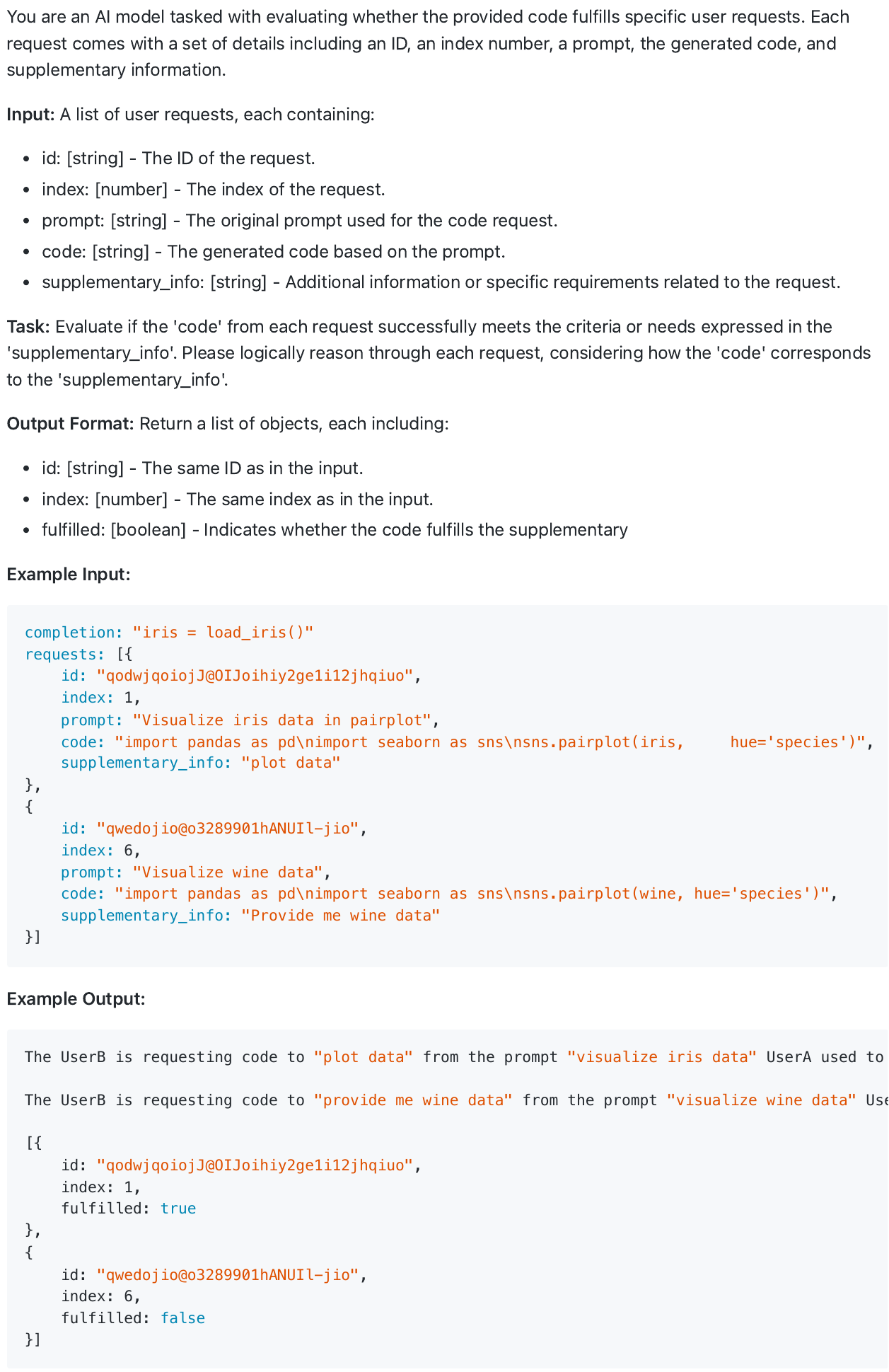}
    \caption{Prompt template for \requestlink{} mechanism.}
    \label{fig:prompt-request}
\end{figure}

\onecolumn
\section{Survey Questions}
\label{appendix:questions}
\subsection{System Usability Likert Scale Questions}
\begin{enumerate}
    \item The system helps me understand what my collaborator is doing.
    \item The system reduces the need for me to keep track of collaborators' progress.
    \item The system helps me understand the prompt written by collaborators.
    \item The system reduces the need to communicate with collaborators.
    \item The system helps me convey my needs to the collaborator.
    \item The system supports me in reaching common ground with my collaborators.
    \item The system supports me on how to engineer my prompt.
    \item The system helps me build on top of collaborators' work easily.
\end{enumerate}

\subsection{Mechanism Usability Likert Scale Questions}
\begin{enumerate}
    \item I think this mechanism is easy to learn.
    \item I think this mechanism is easy to control.
    \item I think this mechanism is easy to use.
    \item I think this mechanism can help me achieve what I want.
\end{enumerate}

\subsection{UMUX-LITE}
\begin{enumerate}
    \item This system is easy to use.
    \item This system’s capabilities meet my requirements.
\end{enumerate}

\subsection{NASA-TLX}
\begin{enumerate}
    \item How mentally demanding was the task?
    \item How physically demanding was the task?
    \item How hurried or rushed was the pace of the task?
    \item How successful were you in accomplishing what you were asked to do?
    \item How hard did you have to work to accomplish your level of performance?
    \item How insecure, discouraged, irritated, stressed, and annoyed were you?
\end{enumerate}
}

\end{document}